\documentclass{aa}  

\usepackage{natbib}
\usepackage{lscape}
\usepackage{graphicx}
\usepackage{txfonts}
\usepackage{adjustbox}
\usepackage{threeparttable}
\usepackage{lscape}
\usepackage{color}
\usepackage{ulem}

   \usepackage[	colorlinks=true, 
   				citecolor=blue,
   				urlcolor=blue,
   				linkcolor=blue,
               pdftex,   
               hyperindex=true,   
               bookmarks=true,   
               bookmarksopen=true,   
               bookmarksnumbered=true]   
               {hyperref}   
\begin{document}

   \title{Tomography of cool giant and supergiant star atmospheres}
   
   \subtitle{II.  Signature of convection in the atmosphere of the red supergiant star  $\mu$ Cep\thanks{Based on observations made with the Mercator Telescope, operated on the island of La Palma by the Flemish Community, at the Spanish Observatorio del Roque de los Muchachos of the Instituto de Astrofísica de Canarias.} }

   \author{K. Kravchenko
          \inst{1,2}, A. Chiavassa \inst{3}, S. Van Eck \inst{2}, A. Jorissen\inst{2}, T. Merle\inst{2}, B. Freytag\inst{4}, 
          \and
         B. Plez \inst{5}
          }

   \institute{European Southern Observatory, Karl-Schwarzschild-Str. 2, 85748 Garching bei M{\"u}nchen, Germany \\
   \email{kateryna.kravchenko@eso.org}
   \and
   Institut d'Astronomie et d'Astrophysique, Universit\'e Libre de Bruxelles,
              CP. 226, Boulevard du Triomphe, 1050 Bruxelles, Belgium
                          \and
             Universit\'e C\^ote d'Azur, Observatoire de la C\^ote d'Azur, CNRS, Lagrange, CS 34229, 06304 Nice Cedex 4, France 
		 \and
             Department of Physics and Astronomy at Uppsala University, Regementsv{\"a}gen 1, Box 516, SE-75120 Uppsala, Sweden      
          \and
             Laboratoire Univers et Particules de Montpellier, Universit{\'e} de Montpellier, CNRS, 34095, Montpellier Cedex 05, France   
            }


 
  \abstract
{Red supergiants are cool massive stars and are the largest and the most luminous stars in the universe. They 
are characterized by irregular or semi-regular photometric variations, the physics of which is not clearly understood. 
}
{The paper aims at deriving the velocity field in the red supergiant star $\mu$ Cep and relating it to the photometric variability with the help of the tomographic method.}
{The tomographic method allows to recover the line-of-sight velocity distribution over the stellar disk and within different optical-depth slices. The method is applied to a series of high-resolution spectra of $\mu$ Cep, and these results are compared to those obtained from 3D radiative-hydrodynamics CO5BOLD simulations of red supergiants. Fluctuations in the velocity field are compared with photometric and spectroscopic variations, the latter being derived from the TiO band strength and serving (at least partly) a proxy of the variations in effective temperature.
}
{The tomographic method  reveals a phase shift between the velocity and spectroscopic/photometric variations. This phase shift results in a hysteresis loop in the temperature -- velocity plane, with a timescale of a few hundred days, similar to the photometric one. The similarity between the hysteresis loop timescale measured in $\mu$ Cep and the timescale of acoustic waves disturbing the convective pattern suggests that such waves play an important role in triggering the hysteresis loops.
}
   {}

\titlerunning{Tomography of cool giant and supergiant star atmospheres}   
\authorrunning{Kravchenko et al.}  
\maketitle

%

\section{Introduction}

Red supergiants (RSGs) represent the late stage of the evolution of massive ($> 8$~M$_{\odot}$) stars before they explode as type~II supernovae. These stars are characterized by high luminosities ($\sim 10^5$~L$_\odot$), low effective temperatures (ranging between 3450 and 4300~K) and low surface gravities (ranging between $\log g = -0.5$ and 0.5); such low gravities result in large radii of the order of  $10^3$~R$_\odot$ \citep{2005ApJ...628..973L,2007A&A...469..671J}.

One of the distinctive properties of RSG stars is their large mass-loss rates \citep[$\dot{M} \sim 10^{-7} - 10^{-4} \, M_{\odot}$/yr;][]{2015A&A...575A..60M} which place them among the major contributors to the chemical enrichment of the interstellar medium. However, the driving mechanism of the mass-loss process in these stars is still poorly understood. \citet{2007A&A...469..671J} proposed that convective motions (which decrease the effective gravity) in the atmosphere of RSG stars combined with the radiative pressure on molecular lines could initiate the mass loss. This scenario was later supported by \citet{2015A&A...575A..50A} who found that the extension of the molecular layers increases with increasing luminosity and with decreasing surface gravity in a sample of RSG stars. An additional  mechanism which may contribute to the mass-loss process is the magnetic field recently detected around RSGs \citep[for example,][]{2010A&A...516L...2A,2017A&A...603A.129T}.

Another relevant property of RSG stars is their photometric variability. \citet{2006MNRAS.372.1721K} analyzed visual light curves of a sample of RSG stars. These light curves are characterized by $V$-band amplitudes between 1 and 4 mag. Various timescales are in general involved in their (pseudo-)periodicity. However, the common behaviour is represented by two main photometric periods: a short one  -- of the order of a  few hundred days, and a long one -- of the order of a few thousand days. The short photometric periods were attributed  to atmospheric pulsations in the fundamental or low-overtone modes, while magnetic activity, binarity or non-radial gravity modes are sometimes invoked to explain the long periods. In addition, according to \citet{1975ApJ...195..137S}, large convective cells in the atmospheres of RSG stars could induce photometric irregularities.

Though more luminous, RSG stars bear some similarities with asymptotic giant branch (AGB) stars of the Mira type. It was shown by \citet{2015A&A...575A..50A} that the (relative) extension of the molecular  layers is comparable in RSG stars and in Mira stars. On the other hand, Mira stars are fundamental-mode pulsators \citep{1999IAUS..191..151W} and are thus characterized by regular photometric variations. Moreover, the $V$-band amplitudes in Mira stars are between 2.5 and 7.0 mag, i.e., in general larger than those of RSG stars. The mass-loss process is also different in Mira stars, where not only convection, but mainly the combination of pulsations and radiation pressure on dust moves the stellar material outwards. Magnetic fields may also contribute to the mass loss \citep{2014A&A...561A..85L}.      

The present paper aims at relating the atmospheric dynamics in the RSG star $\mu$ Cep  to its photometric variability with the help of high-resolution observations and state-of-the-art 3D stellar convection simulations of RSG stars. For this purpose, the recently developed tomographic method, aiming at probing velocity fields at different depths in the stellar atmosphere, will be used.

The paper is structured as follows. Section~\ref{Sect:methodology} introduces the tomographic technique as well as our method to infer the effective temperature. The tomographic method is then applied to the RSG star $\mu$~Cep (Sect.~\ref{Sect: mucep}) as well as to snapshots from 3D radiation-hydrodynamics (RHD) simulations (Sect.~\ref{Sect:3D_simulations}). Section~\ref{Sect:conclusions} summarizes our conclusions.

\section{Methodology}
\label{Sect:methodology}

In order to constrain and characterize the atmospheric motions in $\mu$ Cep, we aim at determining its spatially-resolved velocity field along with its effective temperature as a function of time. Both will then be compared to those provided by 3D RHD CO5BOLD simulations. For this purpose, different techniques will be used. They are described in the following sections.

\subsection{Tomography}

The derivation of the velocity field will be performed using the tomographic technique developed by \citet{2001A&A...379..288A}. The tomographic method aims at recovering the temporarily-resolved line-of-sight velocity distribution as a function of depth in the stellar atmosphere (within different optical-depth slices). The method was further developed and validated on 3D RHD CO5BOLD simulations of RSG stars by \citet{2018A&A...610A..29K}.
In short, the tomographic technique is based on sorting spectral lines according to their formation depths as indicated by the maximum of the contribution function \citep[CF; see][]{2018A&A...610A..29K}. The formation depth is expressed on an optical depth scale computed at the reference wavelength $\lambda = 5000$~\AA. The atmosphere is split into different slices corresponding to specific optical depth ranges. For each slice, a spectral mask is then constructed which contains the wavelengths of lines forming in the corresponding range of optical depths. For a better wavelength precision, only atomic (and not molecular) lines are kept in masks. The masks are then cross-correlated  with either observed or synthetic stellar spectra. The resulting cross-correlation function (CCF) profile reflects the average shape and radial velocity (RV) of lines forming in a given optical depth range.

\subsubsection{Construction of masks}
\label{Sect:mucep_masks}

A set of 5 tomographic masks was constructed from a 1D model atmosphere computed with the MARCS code \citep{2008A&A...486..951G}. The parameters of the model are $T_{\rm eff} = 3400$~K, $\log g = -0.4$, mass = 5 $M_{\odot}$, microturbulence velocity 2~km/s, and solar metallicity. The selected stellar parameters fall in the parameter range of  RSGs \citep[in terms of temperature and surface gravity;][]{2005ApJ...628..973L} as well as of current 3D simulations \citep[in terms of temperature, surface gravity and mass;][]{2011A&A...535A..22C}. In the following, we apply this set of five masks to the RSG $\mu$~Cep, as well as to a 3D simulation. The properties of the masks, i.e., the optical depth boundaries and number of lines per mask, are shown in Table~\ref{tab:masks}.

\begin{table}
\begin{center}
\begin{threeparttable}
\caption[]{Properties of the tomographic masks.}
\label{tab:masks} 
\begin{tabular}{c c c }
\hline \hline
\noalign{\smallskip}
Mask  & $\rm \log \tau_0$ limits\tnote{*} & number of lines   \\
\noalign{\smallskip}
\hline
\noalign{\smallskip}
C1 & $ -1.0 < \log \tau_0 \leq 0.0 $  & 419 \\
C2 & $ -2.0 < \log \tau_0 \leq -1.0 $ & 1750 \\
C3 & $ -3.0 < \log \tau_0 \leq -2.0 $ & 1199 \\
C4 & $ -4.0 < \log \tau_0 \leq -3.0 $ & 433 \\
C5 & $ -4.6 < \log \tau_0 \leq -4.0 $           & 378 \\
\noalign{\smallskip}
\hline
\end{tabular}
\begin{tablenotes}
\item [*] $\tau_0$ is the reference optical depth computed at $\lambda~=~5000$~\AA. 
\end{tablenotes}
\end{threeparttable}
\end{center}
\end{table}

This "new" set of masks is different from the "old" one used by \citet{2018A&A...610A..29K} in terms of number of masks and corresponding optical-depth ranges. The "old" set of masks was built from a 1D MARCS model atmosphere with the same stellar parameters as the one used in the present work but with a different atmospheric extension. The atmospheric extension may be expressed by the Rosseland optical depth $\tau_{\rm Ross}$ (i.e., the optical depth derived from the Rosseland mean opacity). In general, MARCS model atmospheres are computed between $\tau_{\rm Ross} = 10^{-5}$ and 100. In our case, the constructed model extends further to $\tau_{\rm Ross} = 10^{-6}$.

\subsubsection{CCF analysis}
\label{Sect:methodology_RV}

The tomographic masks obtained as described in the previous section will be cross-correlated with $\mu$ Cep and 3D snapshot spectra in Sects.~\ref{Sect:mucep_hysteresis} and \ref{3D_results}, respectively. The cross-correlation is a powerful technique which allows to sum the information from hundreds of spectral lines (which share properties defined during the masks construction) into an average profile, thus increasing the SNR. The quantity
\begin{equation}
p(\lambda) = \frac{1-\rm{CCF}(\lambda)}{\int_0^{\infty} [1-\rm{CCF}(\lambda)] \, \rm{d} \lambda}
\end{equation}
may be interpreted as a line probability distribution \citep{2013EAS....60...85L}. 
Thus, the first (raw) moment of $p(\lambda)$ (hereafter $M_1$) allows to characterize the CCF in terms of the mean RV shift.

In the case of an asymmetric (or double-peaked) CCF, one may derive separately the RV for each CCF component. For this purpose, a fit of the CCF with a single (or multiple) Gaussian profile is performed with the help of the DOE tool \citep{2017A&A...608A..95M}. In brief, DOE takes as input the CCF of a given spectrum and returns the number of peaks present in the CCF. To this end, DOE computes the first, second and third derivatives of the CCF by convolving it with, respectively, the first, second and third derivative of a narrow Gaussian kernel. This technique allows us to differentiate the CCFs and smooth them simultaneously, which avoid the numerical noise that may appear when one does discrete differentiation. Then, DOE looks for ascending zeros of the third derivative, i.e., inflection points (or local minima of the second derivative), to disentangle the CCF  components. The CCF is then fitted by a multi-Gaussian function over a small range around the local maxima and/or inflection points in order to obtain the velocity of the identified peaks.

Finally, for CCFs with  a Gaussian profile, one may assess the full width at half maximum of the CCF as $\rm 2.355 \sqrt{M_2}$, where $ M_2$ is the second (central) moment of $p(\lambda)$, i.e., its variance.

\subsection{Effective temperature determination}
\label{Sect:methodology_Teff}

The derivation of the effective temperature from observed and synthetic spectra will be performed following the method described in \citet{2017A&A...601A..10V}. This method is based on the computation of band-strength indices
\begin{equation}
    B = 1 - \frac{(\lambda_{C_f}-\lambda_{C_i})}{(\lambda_{B_f}-\lambda_{B_i})} \frac{\int_{\lambda_{B_i}}^{\lambda_{B_f}} F_{\lambda} \rm{d} \lambda}{\int_{\lambda_{C_i}}^{\lambda_{C_f}} F_{\lambda} \rm{d} \lambda} \enspace ,
\end{equation}
with
\[
\begin{array}{lp{0.7\linewidth}}
F_{\lambda}       & the observed or synthetic spectrum;\\
\lambda_{C_i}    & lower boundary of the local continuum window;\\
\lambda_{C_f}    & upper boundary of the local continuum window;\\
\lambda_{B_i}    & lower boundary of the band window;\\
\lambda_{B_f}    & upper boundary of the band window. \\
\end{array}
\]

A subset of five TiO bands (see Table~\ref{tab:bands}) was selected from the sample of \citet{2017A&A...601A..10V}. The measured band-strength index is then the mean of single-band indices from Table~\ref{tab:bands}. 

\begin{table}
\begin{center}
\caption[]{Wavelength boundaries (in \AA) of local continuum ($\lambda_{C_{i,f}}$) and TiO band windows ($\lambda_{B_{i,f}}$).}
\label{tab:bands} 
\begin{tabular}{c c c c}
\hline \hline
\noalign{\smallskip}
$\lambda_{B_i}$ & $\lambda_{B_f}$  & $\lambda_{C_i}$ & $\lambda_{C_f}$   \\
\noalign{\smallskip}
\hline
\noalign{\smallskip}
5847.0 & 5869.0 & 5800.0 & 5847.0 \\
6159.0 & 6180.0 & 6067.0 & 6119.0 \\
6187.0 & 6198.0 & 6067.0 & 6119.0 \\
7054.0 & 7069.0 & 7030.0 & 7050.0 \\
7125.0 & 7144.0 & 7030.0 & 7050.0 \\
\noalign{\smallskip}
\hline
\end{tabular}
\end{center}
\end{table}

\begin{figure}
   \centering
   \includegraphics[width=7cm]{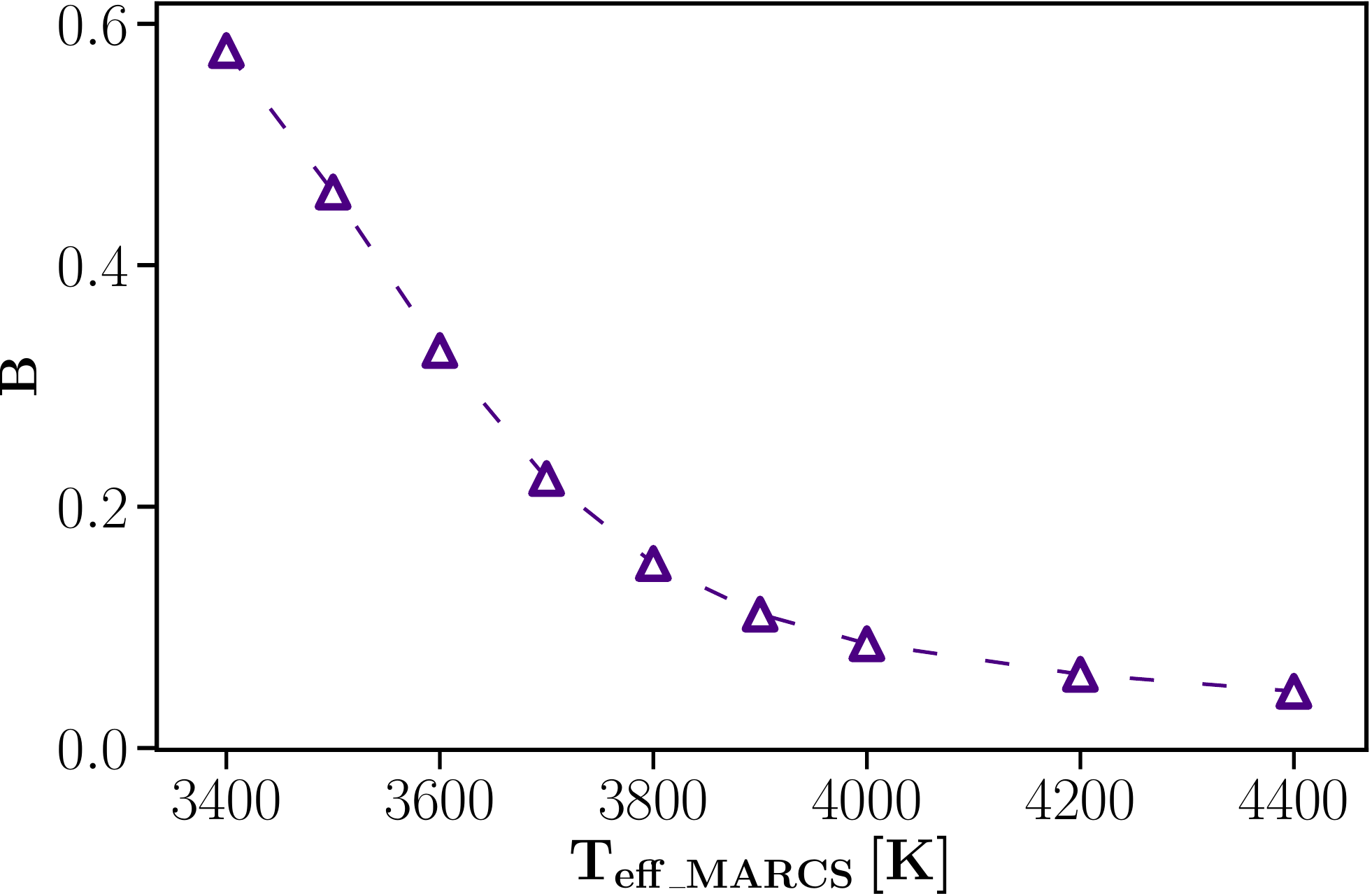}
      \caption{The band-strength index $B$ as a function of  effective temperature as computed from the synthetic spectra of 1D MARCS model atmospheres. }
         \label{band_idx_vs_Teff_MARCS}
   \end{figure}

To derive the effective temperature, we proceed as follows. First, band-strength indices are derived for a grid of 1D MARCS model atmospheres with different effective temperatures in the range 3400 -- 4400~K, $\log g = -0.5$ and solar metallicity. A relation between the effective temperature of 1D model atmospheres and  band-strength indices is thus obtained and shown in Fig.~\ref{band_idx_vs_Teff_MARCS}. Then, band-strength indices are derived for either observed or synthetic spectra and the corresponding effective temperatures are deduced from the above calibration (see Sects.~\ref{Sect:mucep_hysteresis} and \ref{3D_results} for application to $\mu$ Cep and a 3D simulation, respectively).

\section{The red supergiant star $\mu$ Cep}
\label{Sect: mucep}

Our target star $\mu$ Cep was selected from the sample studied by \citet{2007A&A...469..671J} and is one of the largest and brightest RSGs in our Galaxy. Atmospheric parameters of $\mu$ Cep are listed in Table~\ref{tab:mucep_parameters}; they are taken from \citet{2007A&A...469..671J} and \citet{2005ApJ...628..973L}. The photometric periodicity of $\mu$ Cep was analyzed by \citet{2006MNRAS.372.1721K}. The authors computed the power spectrum of $\mu$~Cep visual light curve and deduced two main photometric periods: a short one of about 860~days and a long one of about 4400 days. 

\citet{2008ApJ...685L..75D} have derived the mass-loss rate of $\mu$~Cep from the presence of an extended circumstellar shell at 25 $\mu$m and concluded that it is of the order of a few $10^{-7}$~M$_\odot$~yr$^{-1}$. More recently, \citet{2010A&A...523A..18D}, \citet{2011A&A...526A.156M}, \citet{2016AJ....151...51S}, and \citet{2019MNRAS.485.2417M} derived mass-loss rates of a few $10^{-6}$ M$_\odot$~yr$^{-1}$ for $\mu$~Cep. This value is  much smaller than that of other RSGs, where the mass-loss rate may be orders of magnitude larger  \citep{2006AJ....131..603S}.
VY~CMa for instance has a mass-loss rate of $3 \times 10^{-4}$~M$_\odot$~yr$^{-1}$ \citep{ 2005AJ....129..492H}.

The atmospheric dynamics of $\mu$ Cep was previously studied by \citet{2007A&A...469..671J} by applying the tomographic masks of \citet{2001A&A...379..288A} to a small sample of ELODIE \citep{ELODIE} spectra having a spectral resolution of $R \sim 40 000$ and covering a time span of about 400 days. \citet{2007A&A...469..671J} detected strong asymmetries in the resulting CCF profiles, attributed to complex supersonic velocity fields in $\mu$~Cep atmosphere. They observed as well the time-variable upward and downward motions.

\begin{table*}
\begin{center}
\begin{threeparttable}
\caption[]{Fundamental stellar parameters of $\mu$ Cep.}
\label{tab:mucep_parameters} 
\begin{tabular}{c c c }
\hline \hline
\noalign{\smallskip}
  & Levesque et al. (2005) & Josselin \& Plez (2007)   \\
\noalign{\smallskip}
\hline
\noalign{\smallskip}
$T_{\rm eff} \, [K]$           & 3700 $\pm$ 50 & 3750 $\pm$ 20  \\
$\log g$ [c.g.s.]                   & -0.50  & -0.36 \\
$\rm Mass \, [M_{\odot}]$      & 25\tnote{b}  & 25\tnote{b} \\
$\rm Radius \, [R_{\odot}]$    & 1420  & 1259 \\
$\log (L$/L$_{\odot})$ & 5.53 $\pm$ 0.12\tnote{a} & 5.45 $\pm$ 0.4 \\
\noalign{\smallskip}
\hline
\end{tabular}
\begin{tablenotes}
\item [a] Derived from the bolometric magnitude of -9.08 $\pm$ 0.3 mag. 
\item [b] Derived from evolutionary tracks of \citet{2003A&A...404..975M}.
\end{tablenotes}
\end{threeparttable}
\end{center}
\end{table*}

\subsection{Observations}

The long-term monitoring of $\mu$ Cep was performed with the HERMES spectrograph \citep{2011A&A...526A..69R} mounted on the 1.2m Mercator telescope installed at the Roque de los Muchachos Observatory (La Palma, Spain). The spectral resolution of HERMES is about 86 000 for the high-resolution fiber, and spectra cover the range 3800 -- 9000~\AA.  

In total, 95 high-resolution spectra were obtained for $\mu$ Cep between April 2011 and January 2018, corresponding to a time span of about 2500 days (i.e. $\sim7$ years). The spectra have a typical SNR of $\sim100$ at 5000~\AA. All spectra were reduced using the standard HERMES reduction pipeline, as described by \citet{2011A&A...526A..69R}.

\subsection{Results}
\label{Sect:mucep_hysteresis}

The $\mu$ Cep HERMES spectra were cross-correlated with the set of tomographic masks constructed in Sect.~\ref{Sect:mucep_masks}, { and the corresponding RVs were derived as explained in Sect.~\ref{Sect:methodology_RV}: a fit of the CCFs with single or multiple Gaussian function(s) was performed in order to extract RVs separately for each CCF component}. During the cross-correlation process, the spectra were corrected for the Earth motion. 

As a next step, temperatures were derived for all $\mu$ Cep spectra as described in Sect.~\ref{Sect:methodology_Teff}. Our derived  temperatures are consistent with those obtained by \citet{2005ApJ...628..973L} and \citet{2007A&A...469..671J} (see Table~\ref{tab:mucep_parameters}).

\begin{figure*}
   \centering   
   \includegraphics[width=17.cm]{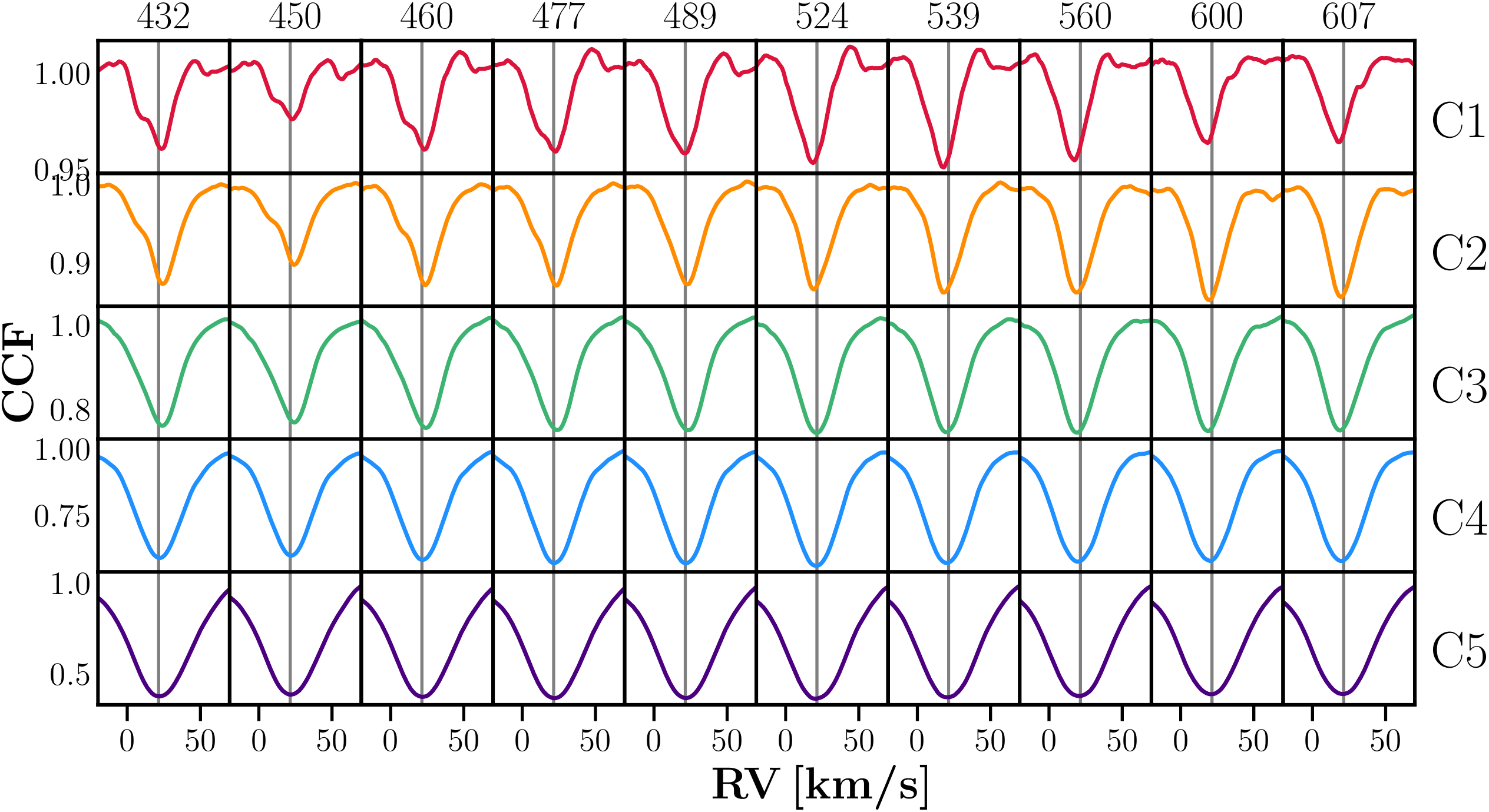}
         \caption{An excerpt of the CCF sequence
          corresponding to the time range between JD~2456432 and JD~2456607 (compare with the light curve of $\mu$~Cep on the bottom panel of Fig.~\ref{mucep_RV}). The last three JD digits are listed on the top of each subpanel.  
          Colors correspond to different masks, mask C1 probing the innermost atmospheric layer while mask C5 probes the outermost layer. Vertical lines at 21.4~km/s indicate the mean radial velocity, computed from all masks and epochs (see text).
         }        
         \label{mucep_hysteresis_ccf}
   \end{figure*}

A small set of CCFs is shown in Fig.~\ref{mucep_hysteresis_ccf}. The CCFs associated with the innermost masks (C1 -- C2) are shallower than those in the outer layers (C4 -- C5), since the former are probing weaker lines (characterized by higher excitation potentials). The CCF widths increase when going from the inner to the outer atmospheric layers. The average CCF width in mask C1 is about 18~km/s while it is about 44~km/s in mask C5.

The CCF profiles show asymmetries in all masks. Moreover, at several epochs, the CCFs in masks C1, C2 and C3 (probing the innermost atmospheric layers) are characterized by an additional blue (or red) component shifted by 10--30~km/s with respect to the main peak. 
Such asymmetries were not observed by \citet{2007A&A...469..671J}, which might be a consequence of the lower resolution of their spectra (twice lower than HERMES), of the more restricted spectral range (3900-6800~\AA, as compared to 3800-8900~\AA\ for HERMES), and/or of the different tomographic masks used.

Line doubling is observed in Mira stars \citep[e.g.,][]{2001A&A...379..288A,2001A&A...379..305A} and is associated with the passage of a shock wave through the atmosphere \citep[known as the Schwarzschild scenario;][]{Schwarzschild}.  
According to this scenario, when the shock front is located far below the line-forming region, the photosphere contains only falling material, so that the spectral lines are red-shifted. Then, as the shock wave approaches the photosphere, a blue component appears and strengthens, associated with the rising matter.  When the shock wave passes through the photosphere, the line is split in two components. Finally, when the shock wave has passed through the photosphere, all material is rising, and the line is  now fully blue-shifted. \citet{2000A&A...362..655A} showed that, in Mira stars, the tomographic method clearly unravels the Schwarzschild scenario as a line-doubling (or shock) front progressing towards external layers as time passes. In other words, in Mira stars, the combined temporal and spatial variations of the CCFs clearly reveal the upward motion of the shock front \citep[see as well Figs. 5 and 6 of][]{2016ASSL..439..137J}. 
In contrast, the temporal and spatial evolution of CCFs in Fig.~\ref{mucep_hysteresis_ccf} does not follow the Schwarzschild scenario.

\begin{figure}
   \centering
   \includegraphics[width=9cm]{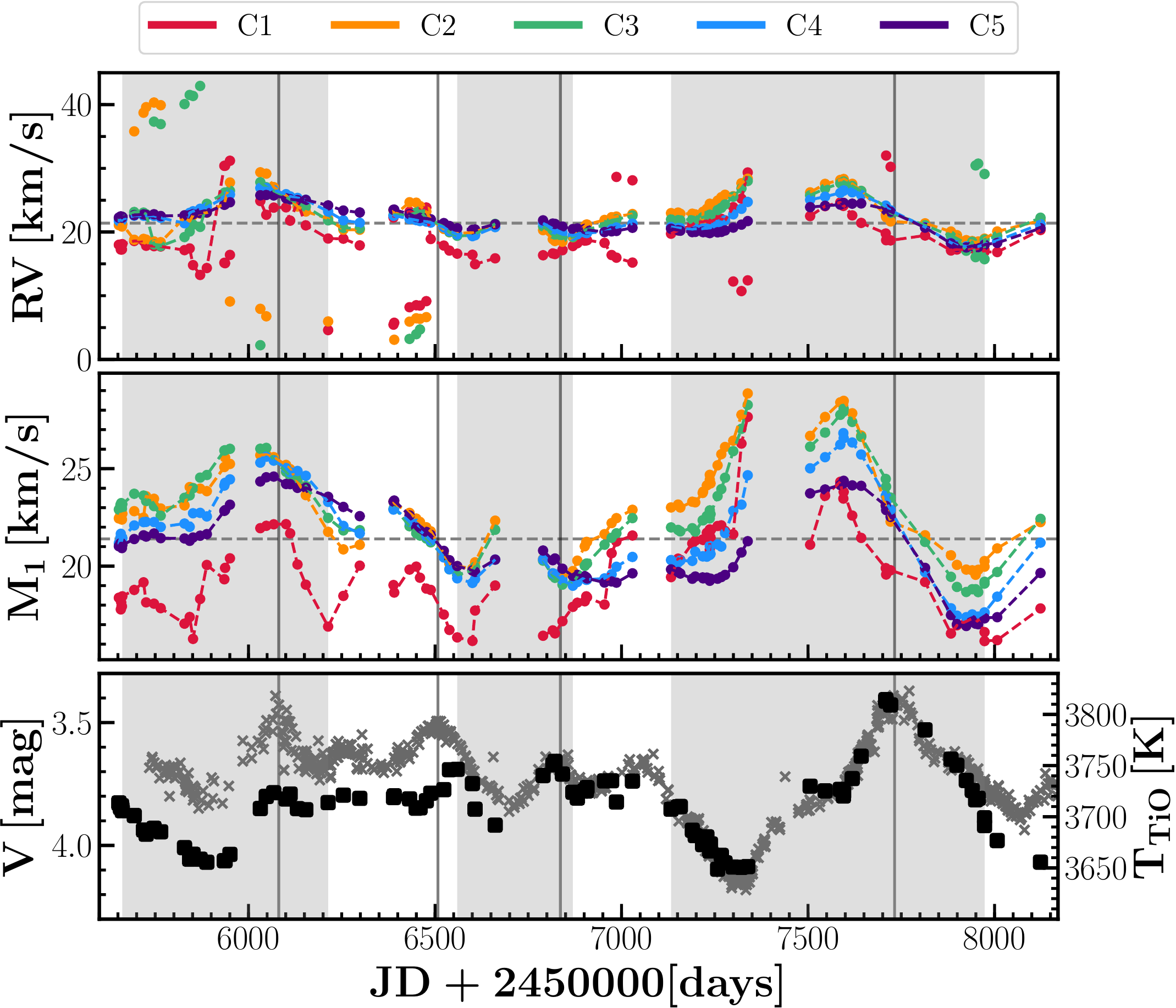}  
      \caption{\textit{Top panel}: the RVs derived from the fit of $\mu$ Cep CCFs with Gaussian functions. RVs from the main CCF component are connected. Only CCF components with depths $\leq 2\sigma$, where $\sigma$ is the standard deviation in the flat part of the CCFs (measuring the "correlation noise"), are kept. Colors correspond to different masks. 
      \textit{Middle panel:} the $M_1$ velocity of CCFs in all masks. The color coding is the same as in the top panel.       
      \textit{Bottom panel:} the AAVSO visual light curve (grey crosses). Black squares correspond to TiO-band temperatures. 
      Vertical lines in all panels indicate times of maximum light and reveal the phase shift between the light (or temperature) and RV variations, the light maximum occurring after the maximum velocity. Horizontal lines in top and middle panels indicate the mean RV. Grey areas 
      define three pseudo light cycles which correspond to the hysteresis loops in Fig.~\ref{mucep_hysteresis}.
              }
         \label{mucep_RV}
   \end{figure}

In masks C1 and C2, at least three velocity components are observed: the main CCF component is always located around 10 - 30 km/s, whereas additional CCF components appear at various variability phases either at 30 - 45 km/s or at 0 - 10 km/s (see also top panel of Fig.~\ref{mucep_RV}). Moreover, the additional CCF component remains weak and never reaches the same contrast as the main CCF peak. 

\begin{figure}
   \centering
	 \includegraphics[width=9cm]{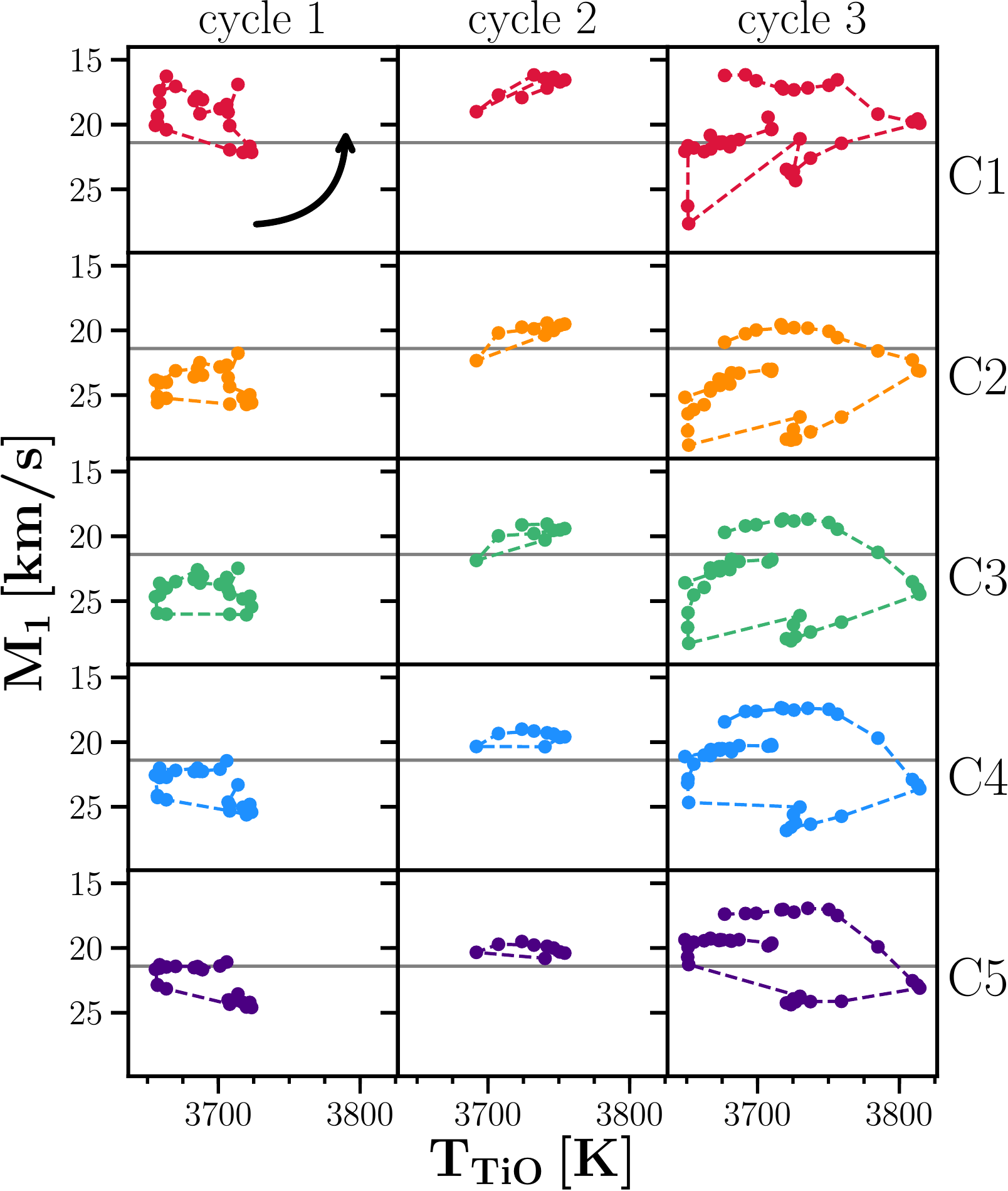} 

      \caption{\textit{From left to right:} Hysteresis loops between TiO-band temperature and $M_1$ velocity for three pseudo light cycles of $\mu$ Cep (identified as shaded areas in Fig.~\ref{mucep_RV}). \textit{From top to bottom:} $M_1$ velocity measured in different masks, from C1 probing the innermost atmospheric layer, to C5 probing the outermost. Horizontal lines in all panels indicate the mean RV from all masks. The arrow indicates the direction of evolution along the hysteresis loops.
      } 
         \label{mucep_hysteresis}
   \end{figure}

Since the center-of-mass (CoM) velocity of $\mu$ Cep is unknown, the distinction between rising and falling material is uncertain. As a proxy of the CoM velocity, we adopt the mean RV of 21.4 km/s (an average over all masks and all epochs; it is represented by the vertical lines in Fig.~\ref{mucep_hysteresis_ccf} and the horizontal lines in Figs.~\ref{mucep_RV}~and~\ref{mucep_hysteresis}). To evaluate the mean RV, the (first-moment) $M_1$ velocity was used (see Sect.~\ref{Sect:methodology_RV}).

The intensity of a CCF component is directly related to the size of the corresponding emitting surface on the star, as shown in Fig.~\ref{maps_for_loops}: for instance, if a large fraction of the stellar surface is covered by rising material, a strong blue-shifted peak will be observed in the CCF.  Thus, asymmetries in the  CCFs hint at the presence of a few granules (or alternatively, at the presence of non-spherically-symmetric shock waves)  in the atmosphere of $\mu$~Cep {which, in any case, behaves differently than the atmospheres of Mira stars.}

Fig.~\ref{mucep_RV} compares the visual light curve extracted from the AAVSO (American Association of Variable Star Observers) database \citep{AAVSO} to {Gaussian-fit or $M_1$ velocities} in the different masks as well as to the TiO-band temperatures.
Fig.~\ref{mucep_RV} reveals that:
\begin{itemize}
\item the TiO-band temperature variation is mostly in phase with the light variation. This agreement, in turn, confirms the reliability of our temperature-derivation method (see Sect.~\ref{Sect:methodology_Teff});

\item the maximum peak-to-peak amplitude of the $M_1$ velocity variations (middle panel of Fig.~\ref{mucep_RV}) is similar for all masks and amounts to about 10~km/s. The corresponding standard deviations lie in the range 2.0 -- 2.5~km/s. This situation contrasts with the results of the 3D RHD simulations discussed in Sect.~\ref{3D_results} where the $M_1$ velocity amplitudes are highest in the outermost mask. An interesting result is that the RV amplitudes in $\mu$ Cep are smaller than those  typical of Mira stars, by a factor of two at least \citep{2001A&A...379..305A};

\item there is a tendency for blue-shifted CCF components to appear in the innermost masks when the stellar brightness increases (top panel of Fig.~\ref{mucep_RV});

\item the RVs in all masks vary with the same periodicity as the light (and temperature), but with a phase shift: the maxima in the light curve occur at later epochs than the velocity maxima. 
The phase shift\footnote{Phase shifts are computed by adopting as unit period the duration of the considered pseudo light cycle (see text for the definition of the pseudo light cycle).} between the $M_1$ curve and the light curve was derived by cross-correlating them for two pseudo light cycles in Fig.~\ref{mucep_RV}.
A pseudo light cycle is defined as one oscillation of the light curve (starting at an arbitrary epoch).
Three such pseudo cycles are identified by the grey shaded areas in Fig.~\ref{mucep_RV}. 
The derived phase shifts for the first and third are similar and amount to about 0.15 for mask C5 and 0.20 for mask C2. Thus, the phase lag decreases towards the outer atmospheric layers.

\end{itemize}

A similar correlation between RV and light variations exists for Mira stars  \citep[e.g., R~CMi; see Fig.~5 of][]{2013EAS....60...85L} and is characterized by a phase shift of about 0.15--0.2. Interestingly,  in their tests of dynamical DARWIN models for M-type AGB stars, \citet{2017A&A...606A...6L} found that a phase shift of 0.2 between  the luminosity and radius maxima has to be introduced in order to reproduce the characteristic RV curves of Mira stars.

Such a phase lag was also detected for the RSG star Betelgeuse by \citet{2008AJ....135.1450G}. It may be represented as a hysteresis loop between the line-depth ratio (\ion{V}{i}~6251.83~\AA~$/$~\ion{Fe}{i}~6252.57~\AA) and the mean core velocity of \ion{V}{i}, \ion{Fe}{i} and \ion{Ti}{i} (6261.11~\AA) spectral lines. The line-depth ratio was considered to be a good temperature indicator and was varying mostly in phase with the light curve. According to \citet{2008AJ....135.1450G}, the hysteresis loop illustrates the convective turn-over of the material in the stellar atmosphere: first, the rising hot matter reaches upper atmospheric layers, then temperature drops as the matter moves horizontally and finally matter falls and cools down. However, the above explanation cannot account for observable time-dependent effects if it relates to purely stationary convection. Therefore, in Sects.~\ref{Sect:3D_maps_for_loops} -- \ref{convective_lifetime} below, we will present a more detailed discussion of the possible origin of the hysteresis loop, searching for mechanisms introducing  non-stationary effects in convection.

Fig.~\ref{mucep_hysteresis} displays the $M_1$ velocity as a function of the TiO-band  temperature for the three pseudo light cycles  of $\mu$ Cep defined above and depicted as shaded areas in Fig.~\ref{mucep_RV}. The hysteresis loops of $\mu$ Cep in Fig.~\ref{mucep_hysteresis} turn counter-clockwise as do those of  Betelgeuse   \citep{2008AJ....135.1450G}. Moreover, thanks to the tomographic method, the hysteresis loops in $\mu$ Cep are now spatially resolved, in the sense that their properties are now probed as a function of depths in the atmosphere. 

The detailed inspection of Fig.~\ref{mucep_hysteresis} and Table~\ref{tab:mucep_loops} reveals that:
\begin{itemize}

\item the hysteresis loops of $\mu$ Cep are characterized by different timescales, as do the corresponding light cycles;

\item there is a  tendency for the RV range of the hysteresis loops of the first and third pseudo-cycles to decrease outwards, going from mask C2 to mask C5 (mask C1 is rather noisy and does not follow this trend). 

\end{itemize}

\begin{table*}[h!]
\begin{center}
\caption[]{Properties of the hysteresis loops of $\mu$ Cep (Sect.~\ref{Sect: mucep}), Betelgeuse \citep{2008AJ....135.1450G}, and the 3D simulation (Sect.~\ref{Sect:3D_simulations}). }
\label{tab:mucep_loops} 
\begin{tabular}{c c c c c c}
\hline \hline
\noalign{\smallskip}
  & mask & RV range  & {$(RV_{\rm min}+RV_{\rm max})/2$} & $T$ range  & Timescale\\
  & & (km/s)& (km/s) & (K) & (d)  \\
  
\noalign{\smallskip}
\hline
\noalign{\smallskip}

       & C1 & 5.9 & 19.2 & & \\
$\mu$ Cep       & C2 & 4.0 & 24.1 & & \\
cycle 1 & C3 & 3.6 & 24.3 & 68 &  552 \\
       & C4 & 4.2 & 23.5 & & \\
       & C5 & 3.5 & 22.8 & & \\
\hline \\        
       & C1 & 2.8 & 17.6 & & \\
$\mu$ Cep       & C2 & 2.9 & 20.9 & & \\       
cycle 2 & C3 & 2.8 & 20.5 & 62 & 310 \\
       & C4 & 1.4 & 19.7 & & \\
       & C5 & 1.3 & 20.3 & & \\
\hline \\       
       & C1 & 11.5 & 22.1 & & \\
$\mu$ Cep       & C2 & 9.3 & 24.2 & & \\
cycle 3 & C3 & 9.6 & 23.5 & 166 & 840 \\
       & C4 & 9.5 & 22.1 & & \\
       & C5 & 7.4 & 20.7 & & \\
\hline \\
Betelgeuse &  & 7-8 & & $\sim$ 100 & $\sim$ 400 \\   
\hline \\  
       & C1 & 2.4 & -0.6 & & \\
       & C2 & 3.1 & 1.1 & & \\
3D     & C3 & 6.5 & 3.4 & 206 & 324 \\
       & C4 & 8.5 & 4.0 & & \\
       & C5 & 9.7 & 5.4 & & \\      
\noalign{\smallskip}
\hline
\end{tabular}
\end{center}
\end{table*}

The characteristic timescale of the third pseudo light cycle (Table~\ref{tab:mucep_loops}) closely matches the short photometric period of $\mu$~Cep \citep[860 days;][]{2006MNRAS.372.1721K}. The timescales of the hysteresis loops for the first and second cycles are shorter by factors of 1.5 and 2.7, respectively. A similar match between the timescales of the hysteresis loops and the light variations was reported  by \citet{2008AJ....135.1450G} for Betelgeuse. The light curve of Betelgeuse is rather regular with a period of about 400 days \citep{2006MNRAS.372.1721K}. The radial-velocity and temperature ranges of the hysteresis loops of Betelgeuse and $\mu$ Cep  are compared in Table~\ref{tab:mucep_loops} and appear to be similar. This similarity  suggests that the same physical process is at work in these objects to trigger the velocity and photometric variations. In order to identify this process, we compare in the next section the results obtained for $\mu$~Cep with those provided by the 3D RHD CO5BOLD simulation of a RSG star atmosphere.

\begin{table*}
\begin{center}
\caption[]{Fundamental parameters of the 3D simulation used in the present work. }
\label{table:1} 
\begin{threeparttable}
\begin{tabular}{c c c c c c c c c c c}
\hline \hline
\noalign{\smallskip}
  & model & grid & $x_{\rm box}$\tnote{b} & $\Delta t$\tnote{b} & $L$ & $T_{\rm eff}$ & $\log \,g$  & Mass  & Radius & Note \\
  &       & [grid points] & ($R_{\odot}$) & (years) & (L$_{\odot}$) & (K) & (c.g.s.) & [$M_\odot$] & [$R_\odot$] &\\
\noalign{\smallskip}
\hline
\noalign{\smallskip}
  & st35gm04n38  & $401^3$ & 1631 & { 11} & { $41035 \pm 1333$\tnote{a}} & $3414 \pm 17$\tnote{a} & $-0.39 \pm 0.01$\tnote{a} & $5.0$ & $582 \pm 5$\tnote{a} & grey\\
\noalign{\smallskip}
\hline
\end{tabular}
\begin{tablenotes}
\item [a] The effective temperature $T_{\rm eff}$, surface gravity $\log \,g$ and stellar radius of the considered 3D simulation are averaged over spherical shells and epochs; errors are one standard-deviation fluctuations with respect to the time average \citep[see][]{2009A&A...506.1351C,2011A&A...535A..22C}. 
\item [b] $x_{\rm box}$ is the extension of the numerical box, $\Delta t$ is the duration covered by the simulation.
\end{tablenotes}
\end{threeparttable}
\end{center}
\end{table*}

\section{3D radiative-hydrodynamics simulations and detailed radiative transfer}
\label{Sect:3D_simulations}

Following \citet{2018A&A...610A..29K}, we used the same 3D RHD simulation of a RSG star computed with the CO5BOLD code \citep{2012jcoph.231..919f}, 
where the combined compressible hydrodynamics and non-local radiative-transfer equations are solved on a Cartesian grid. The code takes into account molecular opacities; however, the radiation transport is treated in a "gray" approximation and ignores radiation pressure and dust opacities. The model geometry is of the kind "star-in-a-box". It is assumed that solar abundances are appropriate for RSG stars. The 3D simulations are time-dependent, characterized by realistic input physics and reproduce the effects of convection and non-radial waves \citep{2011A&A...535A..22C}. Currently, the 3D simulations do not include a  radiative-driven wind. However, in RSGs, the wind velocity that can sustain the mass loss must be very low just above the photosphere due to the high densities there \citep{2017Natur.548..310O}. Therefore, we do not expect that this missing ingredient will jeopardize the comparison with observations performed in the remainder of this paper.

The basic parameters of the 3D simulation are listed in Table~\ref{table:1}. Due to the limited number of currently available 3D simulations of RSG stars, the parameters of the selected 3D simulation differ from those of $\mu$ Cep, especially in terms of stellar mass. 
The computation of a 3D simulation for a 25~M$_{\odot}$ star would require many more grid points than a simulation of a 5~M$_{\odot}$ star, which is already computationally demanding\footnote{If we would go from a 5~M$_{\odot}$ model to a 25~M$_{\odot}$ model with the same effective temperature and radius, i.e. larger surface gravity, we would have to increase the number of grid points by 125. If we would go from a 5~M$_{\odot}$ model to a 25~M$_{\odot}$ model with the same effective temperature and surface gravity, i.e. larger radius, we would have to increase the number of grid points by 11. Both cases are currently out of reach.}. Nevertheless, this 3D simulation reproduces $\mu$ Cep in terms of surface gravity. Moreover, as will be shown later in this Section, the dynamical picture in a 5~M$_{\odot}$ simulation is very different from that of a lower-mass simulation (representing a pulsating AGB star). The former shows the absence of global shocks and produces photometric and spectroscopic signatures similar to those observed in $\mu$~Cep.

The following consequences are expected  from the mismatch between the stellar parameters of $\mu$ Cep and the 3D simulation. First, there 
could be an impact on the granule size $x_g$ which depends on the pressure scale height \citep[see][and references therein]{2018Natur.553..310P}: 
\begin{equation}
\label{Eq:Hp}
H_P = \frac{\Re T_{\rm eff}}{\mu\; g}    
\end{equation}
where
$g$ is the surface gravity,
$\Re$ is the gas constant, and
$\mu$ is the mean molecular weight.
The pressure scale height differs by only 2.5\% when considering gravity and temperature from either the 3D simulation or from  $\mu$~Cep \citep{2007A&A...469..671J}. Since the granule size $x_g$ is roughly proportional to $H_P$ \citep[see][and references therein]{2018Natur.553..310P}, the same (negligible) relative difference ensues for the granule size. However, this means that the ratio of granule size to stellar radius is too large in the simulations (since $\mu$~Cep radius is much larger than the simulation radius; compare Tables~\ref{tab:mucep_parameters} and \ref{table:1}). Second, since the convective efficiency and velocities depend on the effective temperature, convective velocities may be expected to be slightly higher in $\mu$ Cep than in the 3D simulation (since the latter has a lower effective temperature). Third, opacities may slightly differ between $\mu$~Cep and the 3D simulation. Finally, the mass-loss rate may be different in 5 and 25~M$_\odot$ stars. Fortunately, as we discussed in Sect.~\ref{Sect: mucep}, the mass-loss rate of $\mu$~Cep is not as extreme as that observed in other massive RSGs, so that the mass mismatch between model and actual $\mu$~Cep should not be an issue in that respect.

   \begin{figure}
   \centering
    \includegraphics[width=9cm]{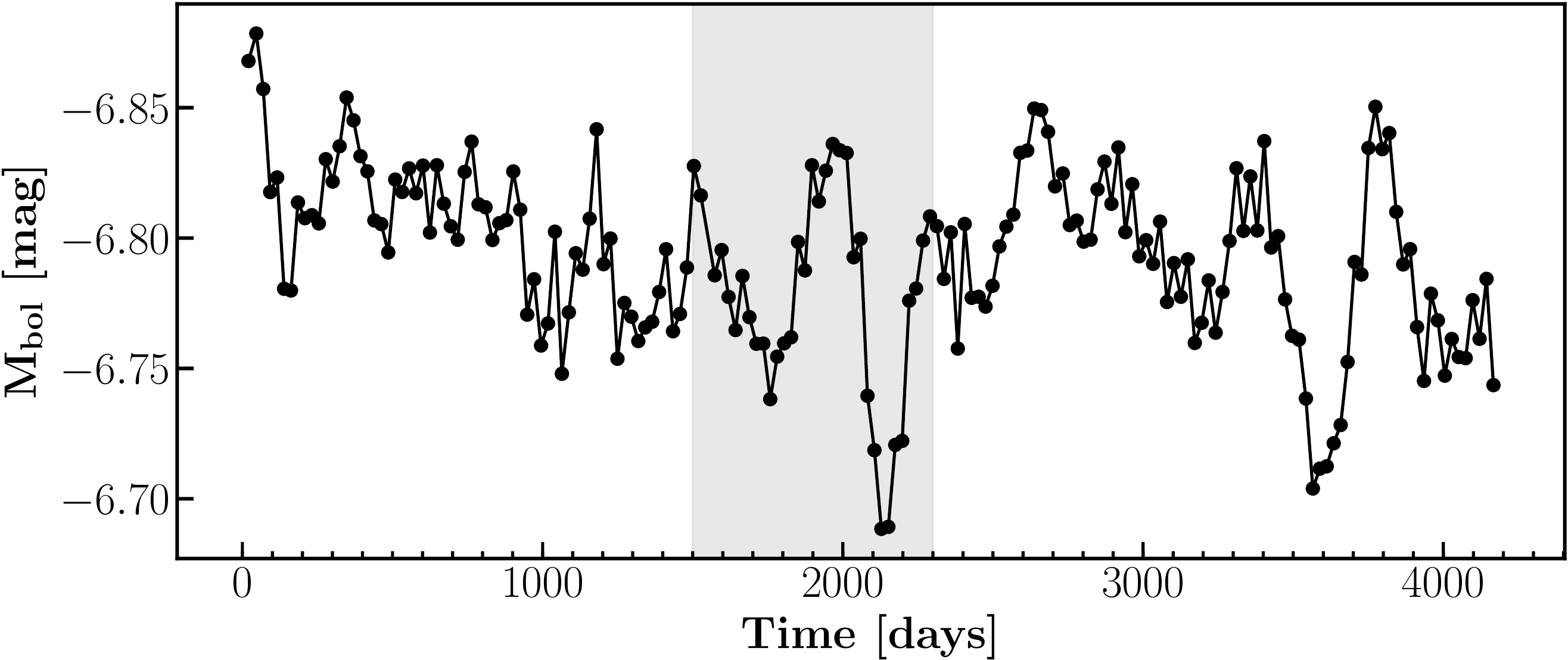}  
      \caption{Bolometric magnitude as a function of time for the 3D RHD simulation. The shaded area corresponds to the subset of snapshots analyzed in Sect.~\ref{3D_results}. }          \label{3D_light_curve}
   \end{figure}

   \begin{figure}
   \centering

    \includegraphics[width=9cm]{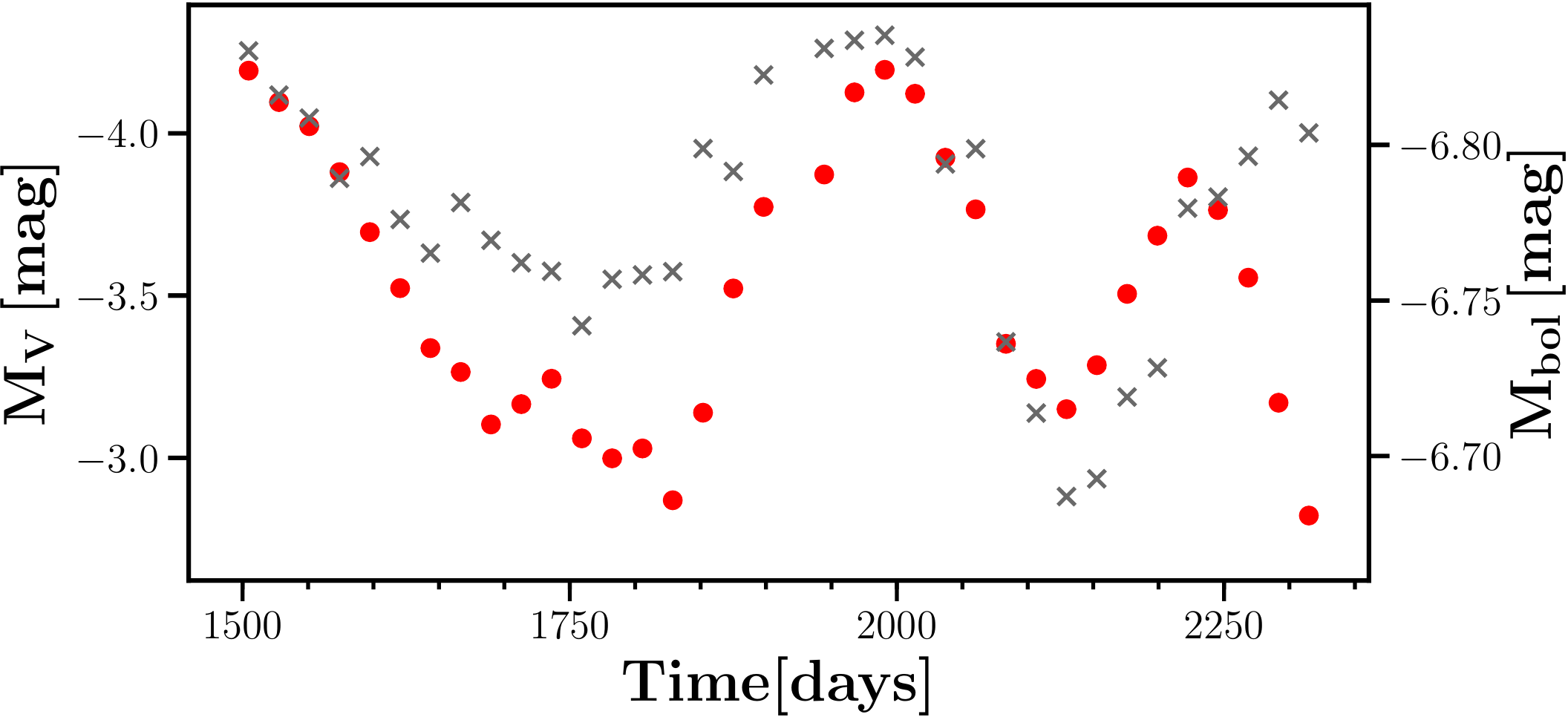} 
      \caption{Absolute visual magnitude (red dots and left scale) and bolometric magnitude (black crosses and right scale) variations during a restricted time span of the 3D RHD simulation (grey-shaded area in Fig.~\ref{3D_light_curve}). The absolute visual magnitude is computed by integrating 3D snapshot spectra in the $V$ band (using the appropriate transmission curve). }
      \label{Fig_3D_BC}  
   \end{figure}

   \begin{figure}
   \centering
   
    \includegraphics[width=9cm]{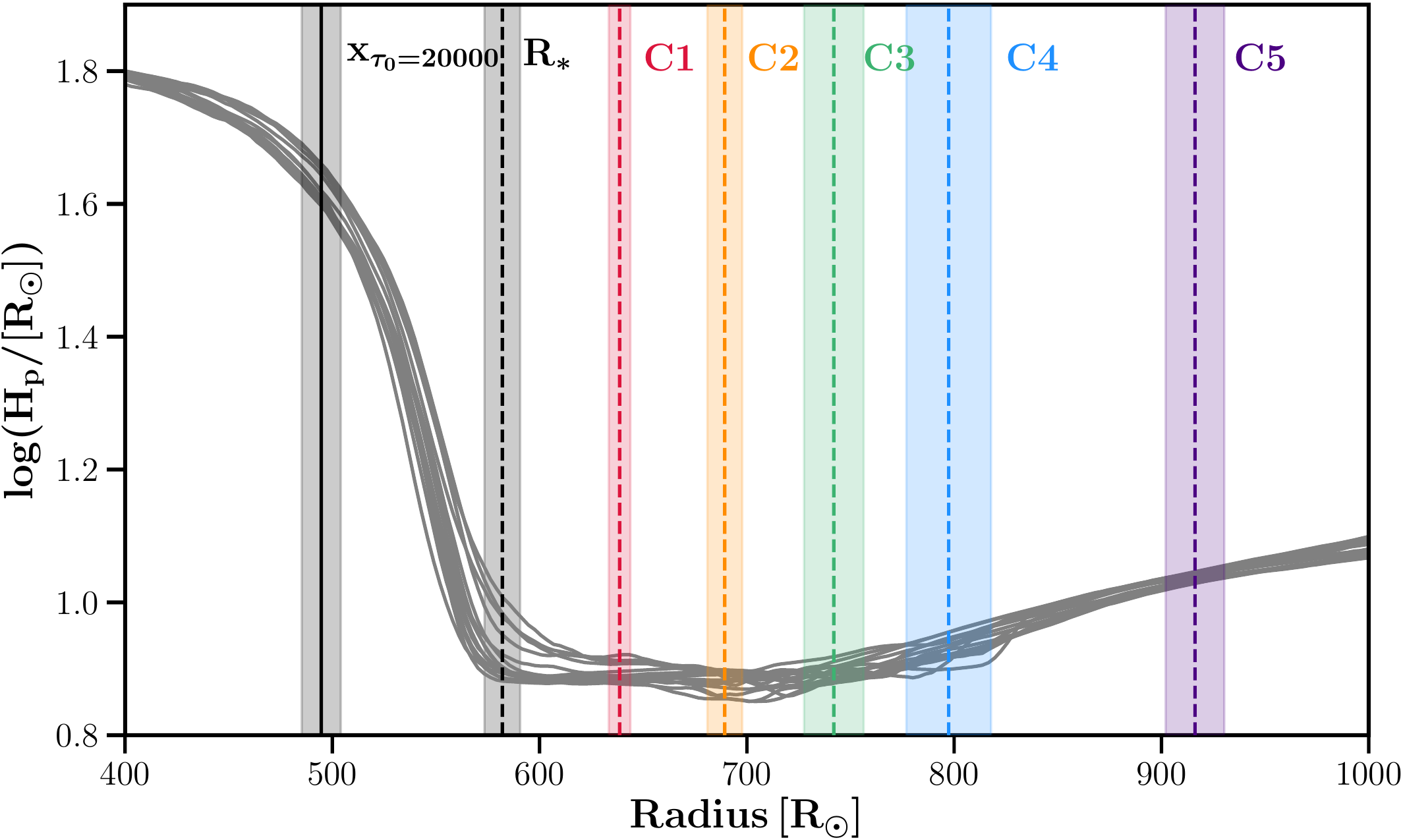}  
      \caption{Pressure scale height ($H_P$) profiles as a function of the shell radius for a set of 3D snapshots. Vertical lines indicate the radial geometrical distances from the stellar center (averaged over a given snapshot and over time) corresponding to the tomographic masks C1-C5, along with the locations corresponding to $\tau_{\rm Ross} = 1$ (denoted as $R_*$) and $\tau_0 = 20000$. The bands are standard deviations of the averages over time.  }
         \label{Hp_vs_masks}
   \end{figure}

   \begin{figure}
   \centering

   \includegraphics[width=7cm]{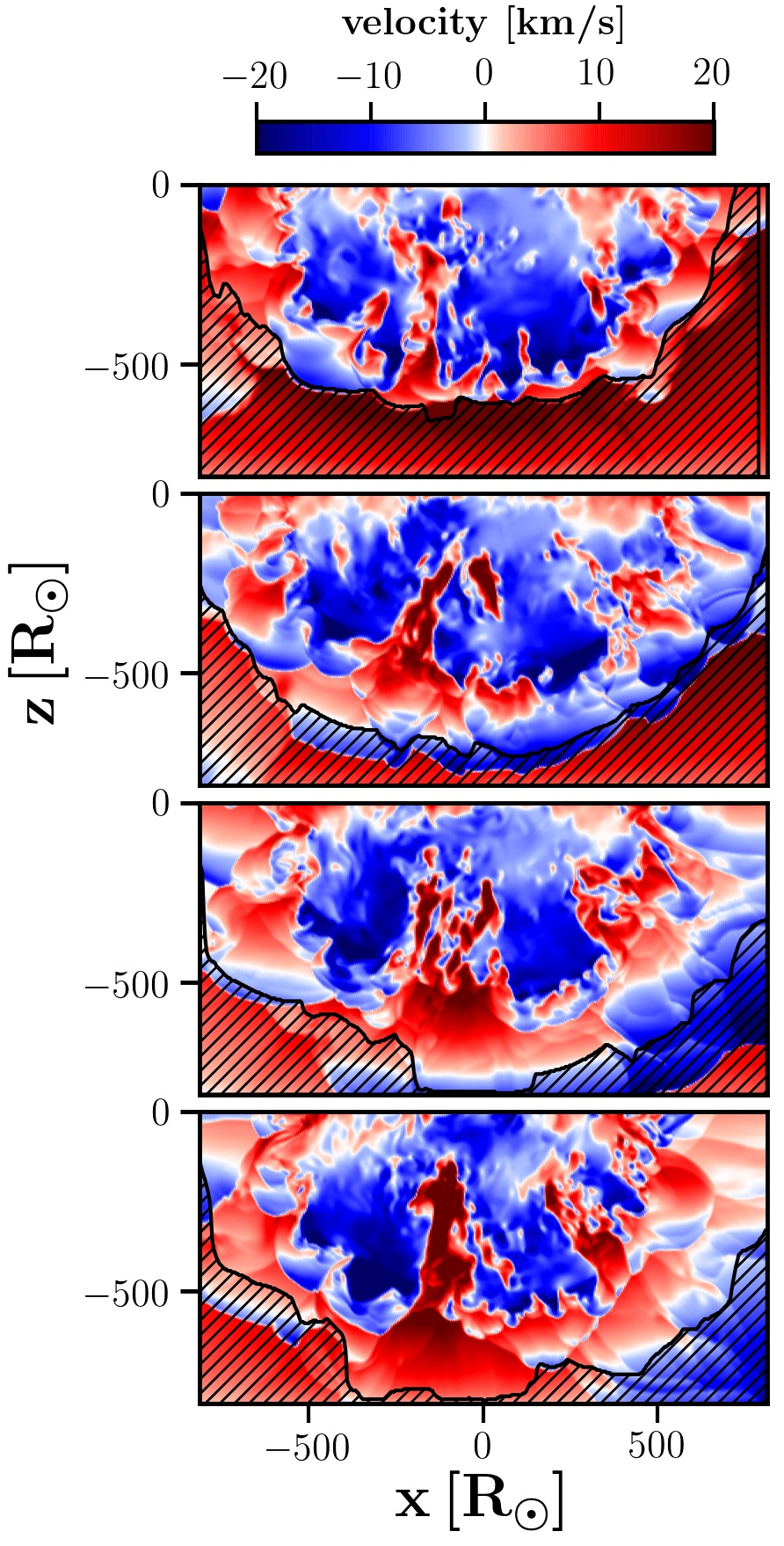}
  
      \caption{The line-of-sight velocity structures for the cut through the center of four different 3D snapshots (for an observer located at the bottom of each panel). Black shaded areas on velocity maps correspond to regions probed by mask C5. }
         \label{appendix_cut_through_the_center_velocity}
   \end{figure}

Fig.~\ref{3D_light_curve} displays the evolution of the bolometric  magnitude over a time span of 11 years covered by the simulations. This synthetic light curve resembles those  typical of RSG stars \citep[for example, see ][and Fig.~\ref{mucep_RV}]{2006MNRAS.372.1721K} in terms of periodicity (of the order of a few hundreds days) and irregular pattern. Because the observed light curve is given in the $V$ filter (Fig.~\ref{mucep_RV}) whereas the 3D RHD simulations generally provide bolometric light curves (Fig.~\ref{3D_light_curve}), Fig.~\ref{Fig_3D_BC} compares the visual and bolometric light curves for a restricted time range of the 3D RHD simulation chosen arbitrarily\footnote{As will be shown later, the results of the paper are independent on the choice of 3D snapshots.} and shown as a grey-shaded area in Fig.~\ref{3D_light_curve}), thus allowing to estimate the bolometric correction. The absolute visual magnitudes were computed by integrating 3D snapshot spectra in the V band using the transmission curve from \citet{1990PASP..102.1181B}. The resulting bolometric corrections are comparable to those of \citet{2005ApJ...628..973L} determined from the MARCS model atmospheres of RSG stars. Furthermore, the visual light curve in Fig.~\ref{Fig_3D_BC} resembles those typical of RSG stars in terms of amplitude ($\sim$~1~mag).

Fig.~\ref{Hp_vs_masks} shows the pressure scale height $H_P$ \citep[spatially averaged over spherical shells as explained in][]{2009A&A...506.1351C} as a function of geometrical depth.
It reveals significantly different values for $H_P$ above and below the stellar radius (corresponding to $\tau_{\rm Ross} = 1$). This will be discussed in Sect.~\ref{surface_vs_deep_convection}. 

The location of each tomographic mask with respect to the stellar radius is shown in Fig.~\ref{Hp_vs_masks} as colored lines. The position of each mask is defined by the average radial distance (from the stellar center) for grid points belonging to the considered mask. The averaging process is performed first over a given snapshot and then over time. The vertical bands in Fig.~\ref{Hp_vs_masks} are one standard deviation fluctuations with respect to the time average. Fig.~\ref{Hp_vs_masks} shows that, as expected, the tomographic masks probe distinct atmospheric layers. It must be noted that mask~C5 is located at just over 900~R$_{\odot}$ from the stellar center, whereas the 3D simulation cube has a half-size of only 800~R$_{\odot}$. Mask~C5 is therefore partially out of the simulation box, and is enclosed within only in its corners, which lie 1150~R$_{\odot}$ away from the stellar center (see Fig.~\ref{appendix_cut_through_the_center_velocity}).

We used the pure-LTE code Optim3D \citep{2009A&A...506.1351C} to compute synthetic spectra for the time span of the 3D simulation corresponding to the gray-shaded area in Fig.~\ref{3D_light_curve}. The code computes the radiative transfer in detail using pre-tabulated extinction coefficients as a function of temperature, density, and wavelength for the solar composition \citep{2009ARA&A..47..481A}. They were constructed with no micro-turbulence broadening, and the temperature and density distributions are optimised to cover the values encountered in the outer layers of the RHD simulations. The code takes into account the Doppler shifts caused by the convective motions. The wavelength range of the computed spectra goes from 3700 to 8900~\AA, and the spectral resolution is $R = 85000$, which mimics the resolution of the HERMES spectrograph (see Sect.~\ref{Sect: mucep}). A few issues associated with the numerical resolution of the 3D simulation were encountered and resolved as explained in Appendix~\ref{problem_with_numerical_resolution}.

\subsection{Results}
\label{3D_results}

   \begin{figure*}
   \centering
   
   \includegraphics[width=18cm]{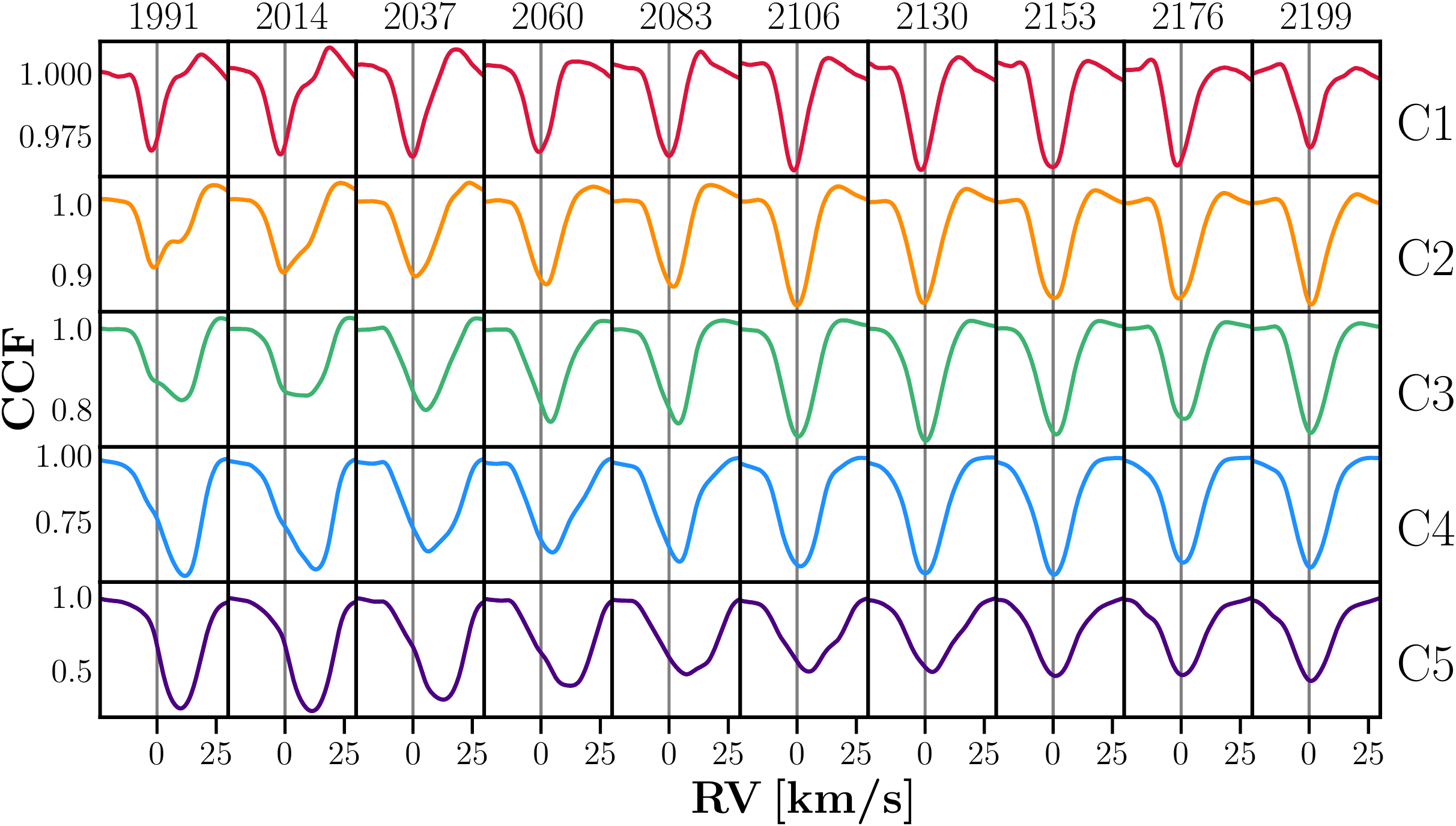}

         \caption{An excerpt of the CCF sequence corresponding to the time range between days 1990 and 2200  of the 3D RHD simulation (see the corresponding light curve in the bottom panel of Fig.~\ref{simulated RVs 2}), for the same masks as in  Fig.~\ref{mucep_hysteresis_ccf}. The day number corresponding to each snapshot is indicated on top of each subpanel. Vertical lines in all panels indicate the 0 km/s velocity (i.e. the CoM velocity of the 3D simulation). 
         }

         \label{3D_hysteresis_CCF}
   \end{figure*}

The synthetic spectra obtained as described in the previous section were cross-correlated with the tomographic masks obtained in Sect.~\ref{Sect:mucep_masks}. An example of resulting CCFs is shown in Fig.~\ref{3D_hysteresis_CCF}. The CCFs appear asymmetric in all masks. Moreover, at several epochs, the CCFs are characterized by an additional blue (or red) component shifted by $\sim 10$~km/s with respect to the main peak. A similar behaviour was observed in $\mu$~Cep CCFs for masks C1--C3 (Fig.~\ref{mucep_hysteresis_ccf}). But unlike in $\mu$ Cep, the simulated CCFs show hints of double-peaked profiles in the outermost masks C4 and C5. 
As for $\mu$ Cep, the simulated CCFs do not follow the Schwarzschild scenario.

Like those of $\mu$ Cep, the simulated CCFs associated with the innermost masks are shallower than those associated with the outer layers. The width of CCF profiles increases going from the inner to the outer atmospheric layers. The average CCF widths in masks C1 and C5 are $\sim 10$~km/s and $\sim 16$~km/s, respectively,  which is a factor of 2 to 2.5 smaller than those of $\mu$~Cep. This indicates that the turbulence velocity fields reached by current 3D simulations are still too low with respect to those observed in real RSG stars. 
This could result from the following shortcomings of current 3D simulations. First, the numerical resolution (i.e. the number of grid points per pressure scale height) of the 3D simulations is still rather low. The size of each grid point is about 4 $R_{\odot}$, which introduces difficulties in resolving the convective structures and the shocks which impact them. Second, the effective temperature of the model used in the 3D simulation is lower than that of $\mu$ Cep (compare Tables~\ref{tab:mucep_parameters} and \ref{table:1}) and, as noted in Sect.~\ref{Sect:3D_simulations}, the amplitude of the turbulent velocity field is related to the effective temperature. Finally, the atmospheric extension of the current 3D simulations is still too low \citep{2015A&A...575A..50A}. Larger and more diluted photospheres would encompass higher Mach numbers.

\begin{figure}
   \centering   

    \includegraphics[width=9.1cm]{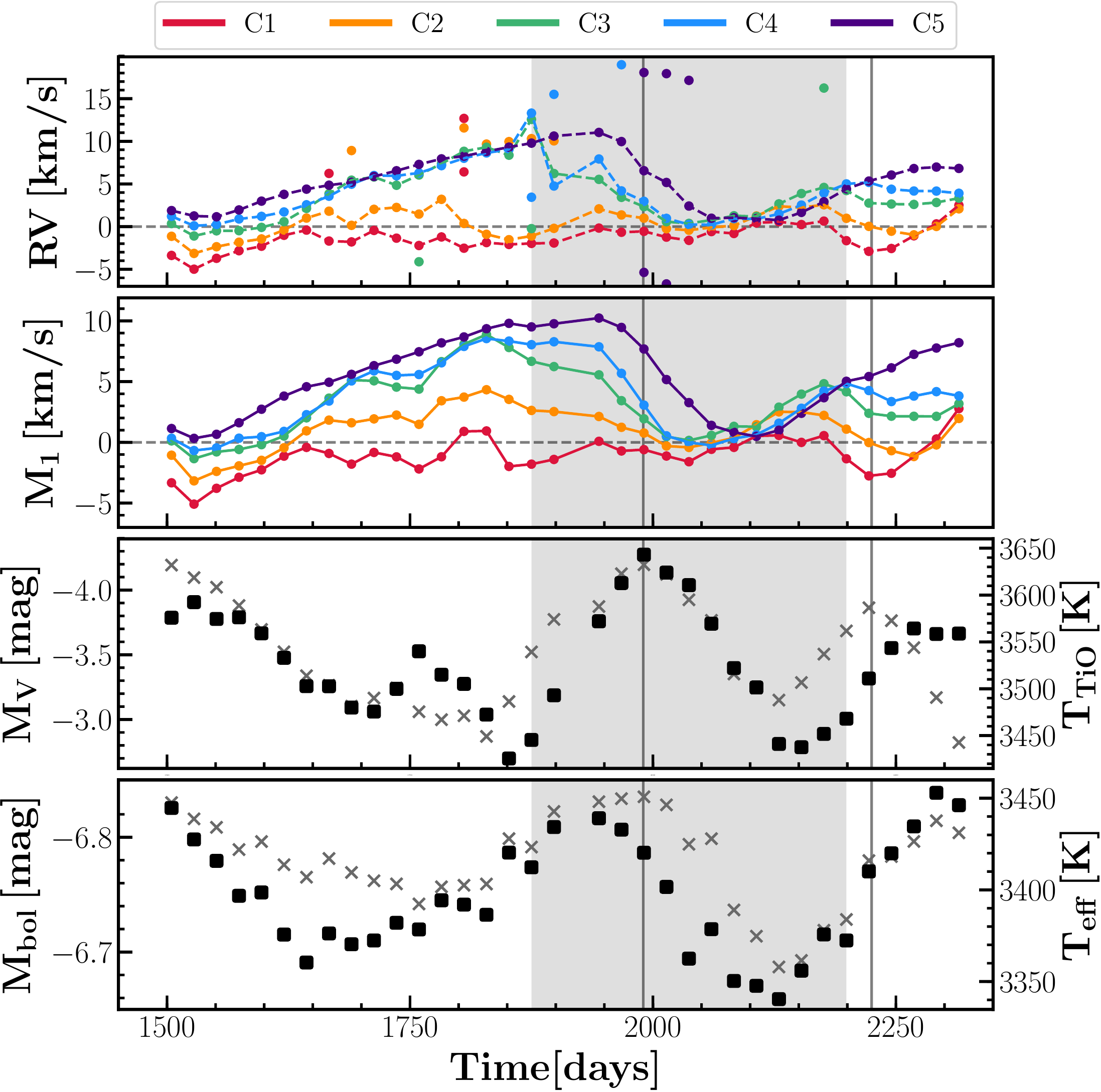} 
          \caption{ Same as Fig.~\ref{mucep_RV}
          for the snapshots of the 3D RHD simulation (except for bottom panel). Bottom panel shows $T_{\rm eff}$ (squares) and bolometric magnitude (crosses) variations for 3D snapshots.  }
         \label{simulated RVs 2}
   \end{figure}

For all simulated CCFs, the corresponding radial velocities were derived as described in Sect.~\ref{Sect:methodology_RV}, and the TiO-band temperatures were computed as explained in Sect.~\ref{Sect:methodology_Teff}. Following \citet{2009A&A...506.1351C}, the effective temperature $T_{\rm eff}$ of a 3D snapshot is defined as the temperature $T$ at radius $R$ for which $L(R)/(4 \pi R^2) = \sigma T^4$ (where $\sigma$ is the Stefan-Boltzmann constant, $L$ is the luminosity). The bottom panel of Fig.~\ref{simulated RVs 2} shows $T_{\rm eff}$ and bolometric magnitude variations for the 3D snapshots. The $T_{\rm eff}$ amplitude of variation explains the $M_{\rm bol}$ amplitude, and they are mostly in phase. The TiO-band temperature variations are twice as large as the $T_{\rm eff}$ variations (third panel of Fig.~\ref{simulated RVs 2}).  The TiO-band temperature follows the variations of $T_{\rm eff}$ but lags behind. The TiO temperature is based on TiO-band strength and is, thus, very sensitive to the upper photospheric temperature stratification. The TiO temperature appears, however, as a good proxy of $T_{\rm eff}$ (and $M_{\rm bol}$, and the stellar variability), with possibly a slight phase shift. It is also the only temperature we can derive in $\mu$~Cep, and use to compare our models with observations.

The evolution of the RVs, $M_1$ velocities, TiO-band temperatures, and visual magnitudes is displayed in Fig.~\ref{simulated RVs 2}. From this comparison, the following conclusions emerge:

\begin{itemize}
\item like in $\mu$ Cep, the temperature variation mimics the light variation (except for the time range between days 1750 and 1900);

\item unlike in $\mu$ Cep, the amplitude of the $M_1$ velocity variations is larger in the outermost masks C3, C4 and C5 ($\sim$ 10 km/s peak-to-peak) than in the innermost masks C1 and C2 ($\sim$ 5 km/s peak-to-peak). In terms of standard deviations of the $M_1$ variations, they increase from  1.5~km/s in mask C1 to 3.2~km/s in mask C5. These values are similar to those observed in $\mu$~Cep (2.0 -- 2.5 km/s);

\item there is no clear correlation between the episodes of line doubling (top panel of Fig.~\ref{simulated RVs 2}) and the light curve. The appearance of a red-shifted secondary peak is more frequent;

\item the $M_1$ velocity  varies with the same periodicity as the light (or  temperature), but with a phase shift of a few tens of days. Similarly to $\mu$~Cep, the maxima of the light curve occur at later epochs than the maxima on the $M_1$ curves. The derived phase shift between the $M_1$ curve and the light curve (for the time span corresponding to the pseudo light cycle in Fig.~\ref{simulated RVs 2}) is 0.31 for mask C3 and 0.19 for mask C5. Thus, as in $\mu$ Cep, the phase lag decreases from mask C3 to mask C5 (due to the irregular behavior of $M_1$ in masks C1 and C2, it was not possible to derive the corresponding phase shifts).

\end{itemize}

\begin{figure}
   \centering

   \includegraphics[width=6cm]{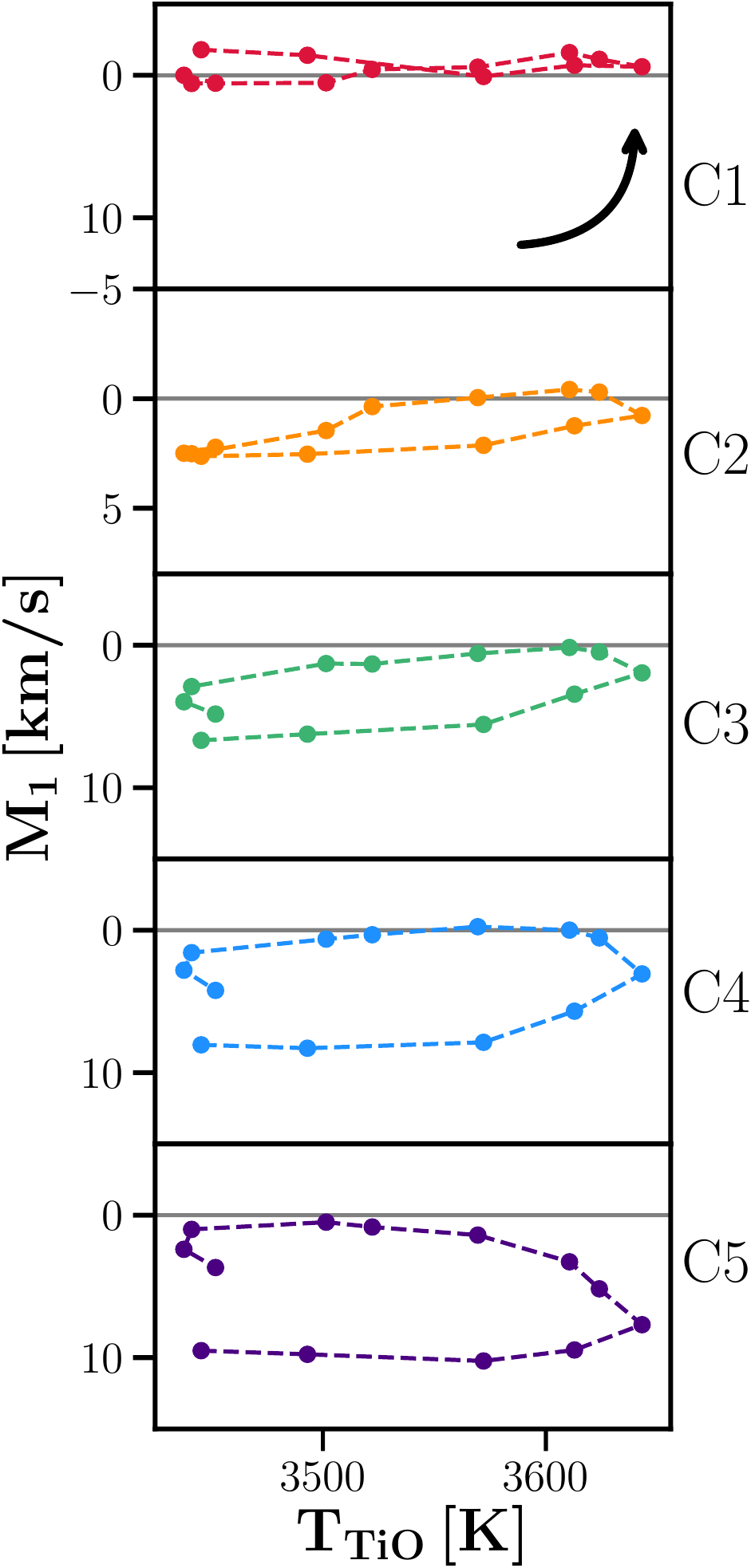}

      \caption{ The hysteresis loop between the TiO-band temperature and the $M_1$ velocity for snapshots from the 3D simulation. Colors correspond to different masks, with mask C1 probing the innermost atmospheric layer, and mask C5 the outermost. Horizontal lines in all panels indicate the CoM velocity of the 3D simulation, i.e. 0 km/s. The arrow indicates the direction of the evolution along the hysteresis loop.
      }
         \label{3D_hysteresis}
   \end{figure}

According to the above results, we conclude that there is a strong similarity between $\mu$ Cep and the 3D RHD simulation, especially the existence of a phase shift between the light (or temperature)  and the velocity variations, which translates into a hysteresis loop (Fig.~\ref{3D_hysteresis}). The hysteresis-like behaviour observed for the 3D RHD simulation  resembles that of $\mu$ Cep (Fig.~\ref{mucep_hysteresis}) and Betelgeuse \citep{2008AJ....135.1450G}. Fig.~\ref{3D_hysteresis} reveals that:

\begin{itemize}
\item unlike in $\mu$ Cep, the hysteresis RV range in the 3D RHD simulation decreases towards the interior (see as well Table~\ref{tab:mucep_loops}). This is a direct consequence of the larger $M_1$ amplitude  in the outermost masks (middle panel of Fig.~\ref{simulated RVs 2}); 

\item since the CoM velocity of the 3D RHD simulation is known (0~km/s), the distinction between rising and falling material is possible. It appears that the upper part of the loop corresponds to stationary matter in all masks, whereas the lower part of the loop is increasingly  redshifted (falling matter) as one considers layers further up. This behaviour is easily accounted for with the help of Figs.~\ref {maps_for_loops} and \ref{Fig:RV_map}. On the upper branch of the hysteresis curve, rising and falling material cover similar areas at the surface, giving rise to a zero radial velocity. On the lower branch of the hysteresis curve, the rising material is still present, but it occupies a smaller surface than the falling material. 

\end{itemize}

For the 3D RHD simulation like for $\mu$ Cep, the RV and temperature ranges as well as the characteristic timescale of the  hysteresis loop were estimated. They are compared to those of $\mu$ Cep and Betelgeuse in Table~\ref{tab:mucep_loops} and all share the same order of magnitude\footnote{Note that in the case of using $T_{\rm eff}$ instead of TiO-band temperature, the temperature range of the hysteresis loop would be $\sim$ 100 K. Nevertheless, it would still be of the same order of magnitude as for $\mu$~Cep and Betelgeuse.}. Thus, understanding the origin of the hysteresis loops from the 3D RHD simulation is crucial for correctly interpreting those observed in actual RSG stars. This is the subject of the next section.

\subsection{Understanding hysteresis loops from the 3D RHD simulation}
\label{Sect:3D_maps_for_loops}

The 3D simulations provide an opportunity to follow the evolution of temperature and velocity along the hysteresis loop, and hence to understand its physical origin. Fig.~\ref{maps_for_loops} displays velocity and temperature maps for different snapshots along the hysteresis loop of mask C4 (Fig.~\ref{3D_hysteresis}). The maps are weighted by the contribution function  of a line contributing to the mask C4 \citep[see][for details]{2018A&A...610A..29K}. The contribution function reveals at which depth (and hence velocity and temperature) in the atmosphere that spectral line forms. 

The velocity maps in Fig.~\ref{maps_for_loops} reveal upward and downward motions of matter extending over large portions of the stellar surface. The relative fraction of upward and downward motions is what distinguishes the upper from the lower part of the hysteresis loop, its top part (zero velocity) being characterized by equal surfaces of rising and falling material. The bottom part of the hysteresis loop occurs, as expected, when the stellar surface is covered mostly by downfalling material. 

On the other hand, the weighted temperature maps in Fig.~\ref{maps_for_loops} can be considered as a proxy to the surface intensity. The maps clearly demonstrate the brightening  of those regions of the stellar surface where matter is about to  rise. These maps make a clear distinction between the left and right parts of the hysteresis loop, the right part being characterized by the presence of high temperatures at several locations on the stellar surface. 

The general trend observed in Fig.~\ref{maps_for_loops}, as described above, confirms that hysteresis loops reflect the  turn-over of material in the stellar atmosphere: the appearance of bright and warm regions at the stellar surface is followed by the rising of material (as seen on the weighted velocity maps) at those same locations, thus accounting for the phase shift observed in Fig.~\ref{simulated RVs 2}.

Figure~\ref{maps_for_loops_NEW} displays the weighted temperature and velocity maps along the hysteresis loop of mask C4 constructed for a different time range of the 3D simulation (i.e. between days 1500 and 1950). The temperature and velocity maps show the same trend as the one observed in Fig.~\ref{maps_for_loops}, thus, confirming that the same physical mechanism takes place in any subset of 3D snapshots.

The appearance and disappearance of warm regions on the weighted temperature maps in  Fig.~\ref{maps_for_loops} are thus responsible for the surface brightness variations. One needs, however, to stress a crucial property of convection in both AGB and RGB stars, as revealed by the 3D simulations  \citep[e.g.,][for AGB stars]{2017A&A...600A.137F}, namely the fact that  the continuum forms {\it above} the top of the convection zone. In both RSG and AGB stars, the deep large-scale convective cells (as displayed on Figs.~\ref{Fig:RV_map} and \ref{maps_for_loops_largetau}) are not directly observable. In RSG stars, we see almost down to the top of the convection zone. The continuum-forming layers sit so close to the top of the convection zone that both move together.\footnote{This is similar to the situation prevailing in the Sun, where the continuum also forms close to the top of the convection zone. Therefore, bright granules and dark intergranular lanes are well visible and have a direct effect on the emergent intensity.} However, the structures we see are non-stationary surface granules,
that are affected by acoustic waves or pulsations.
 
In contrast, in some of the cooler AGB models, we  often see much higher layers, far above the convection zone. Here, the structures can still be shaped by convection, because waves have travelled through the top of the convection zone and thus have been shaped by variations in sound speed, density, and velocity. These stochastic shocks generated and shaped by convection transfer the heat through the atmosphere and, in turn, cause the surface brightness variations.

\subsection{Surface vs. deep convection}
\label{surface_vs_deep_convection}

\begin{figure*}
   \centering
   
   \includegraphics[width=15cm]{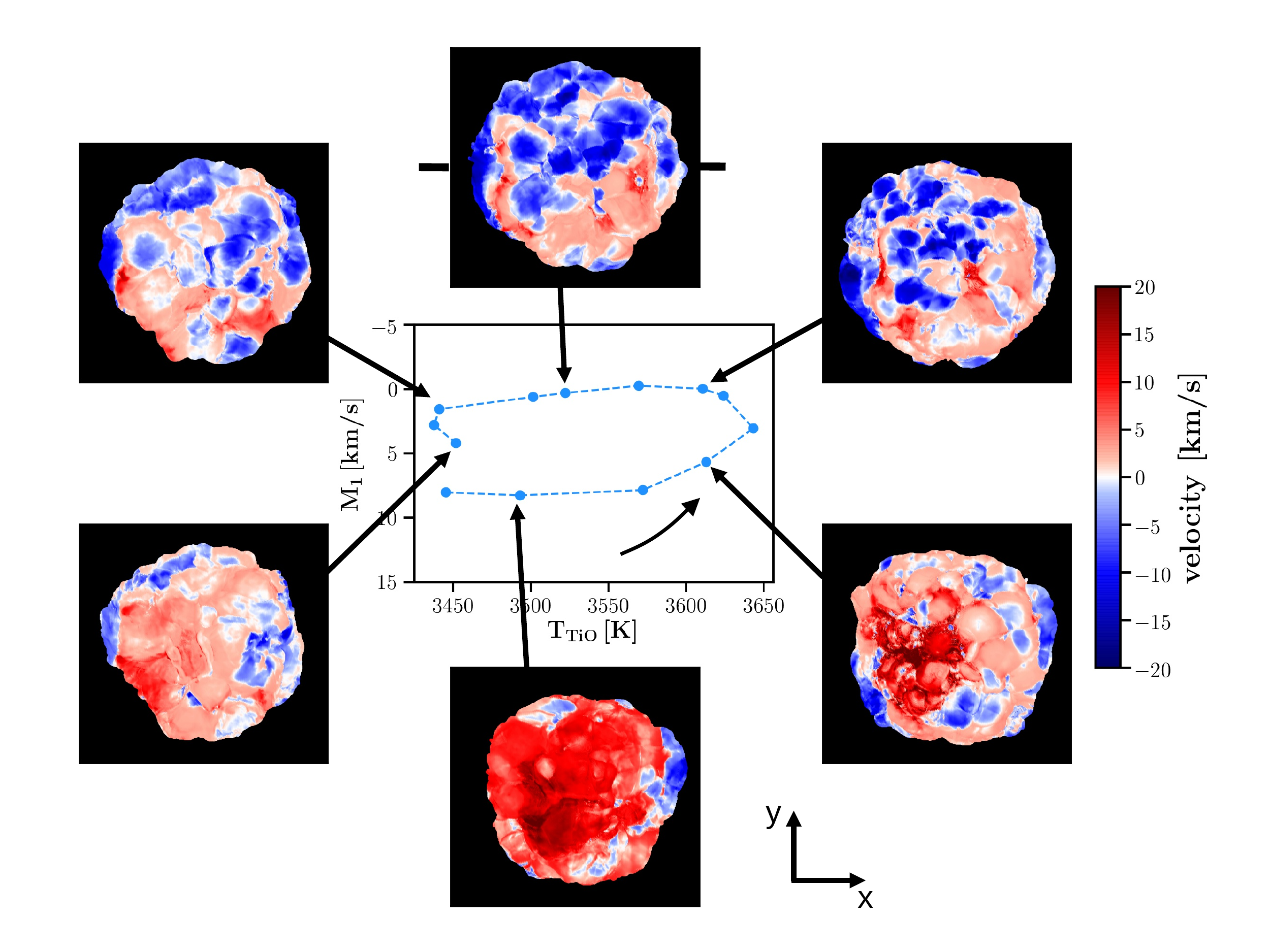}\\
   \includegraphics[width=15cm]{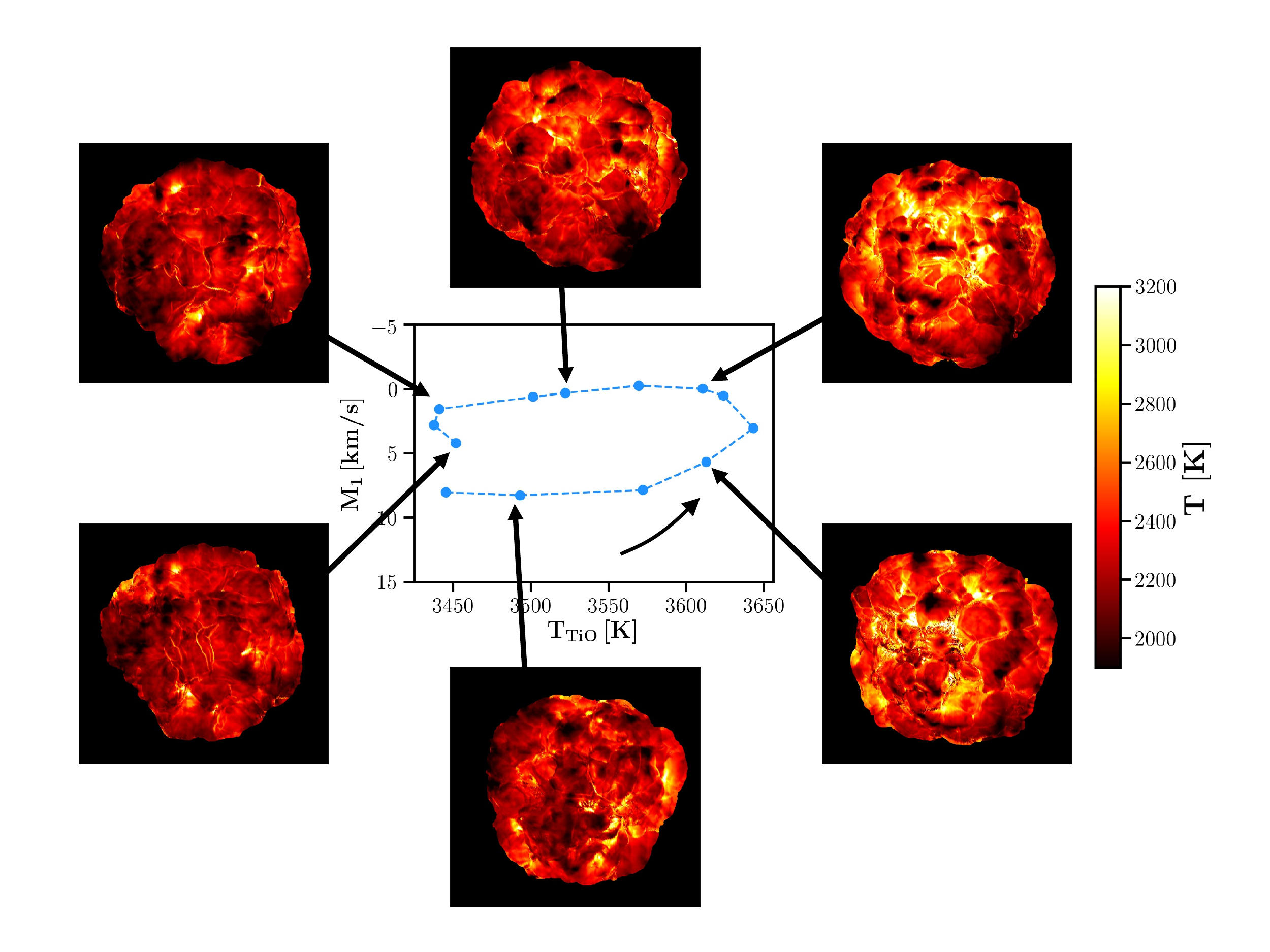}
      \caption{\textit{Top panel:} the line-of-sight velocity   maps (for an observer located in front of the figure) weighted by the CF for snapshots along the hysteresis loop of mask C4. Red color corresponds to falling material, and blue color to rising material. 
      \textit{Bottom panel:} Same as top panel for temperature.
      The arrow indicates the direction of the evolution along the hysteresis loop. The horizontal ticks on the top velocity map show the location of the equatorial plane of 3D snapshots displayed in Fig.~\ref{Fig:RV_map}. 
      }
         \label{maps_for_loops}
   \end{figure*}

\begin{figure*}
   \centering

 \includegraphics[width=15cm]{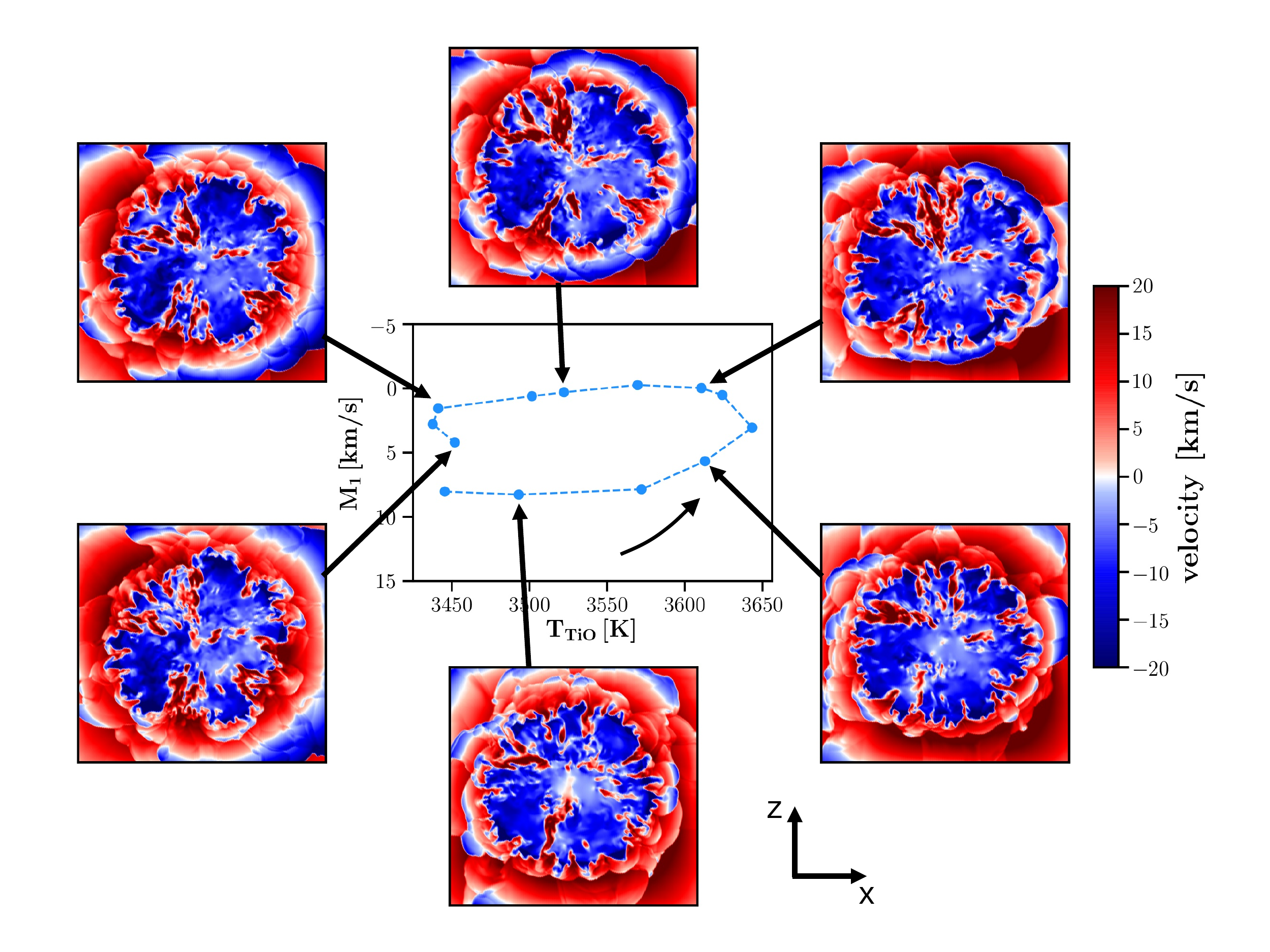}
      \caption{  Velocity maps computed along rays from the center of the star in an equatorial plane for snapshots along the hysteresis loop of mask C4 (the online version of the paper displays an animated version of this figure). 
       This figure displays the large-scale structure of the convection, from the stellar center to the stellar surface. Moreover, it illustrates the appearance of shocks at the stellar surface and their propagation through the atmosphere.
      }
         \label{Fig:RV_map}
   \end{figure*}

\begin{figure*}
   \centering

   \includegraphics[width=15cm]{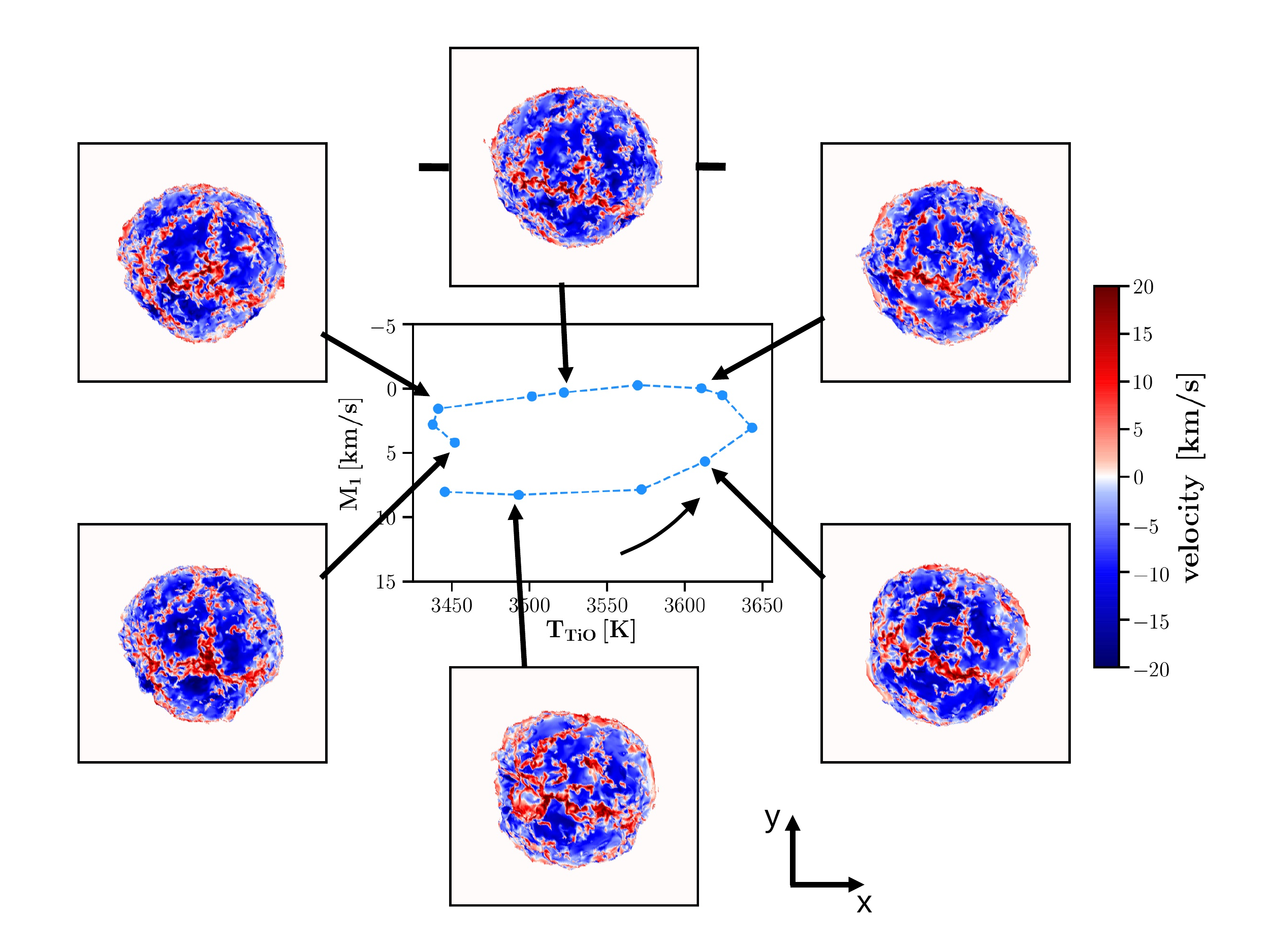}\\
   \includegraphics[width=15cm]{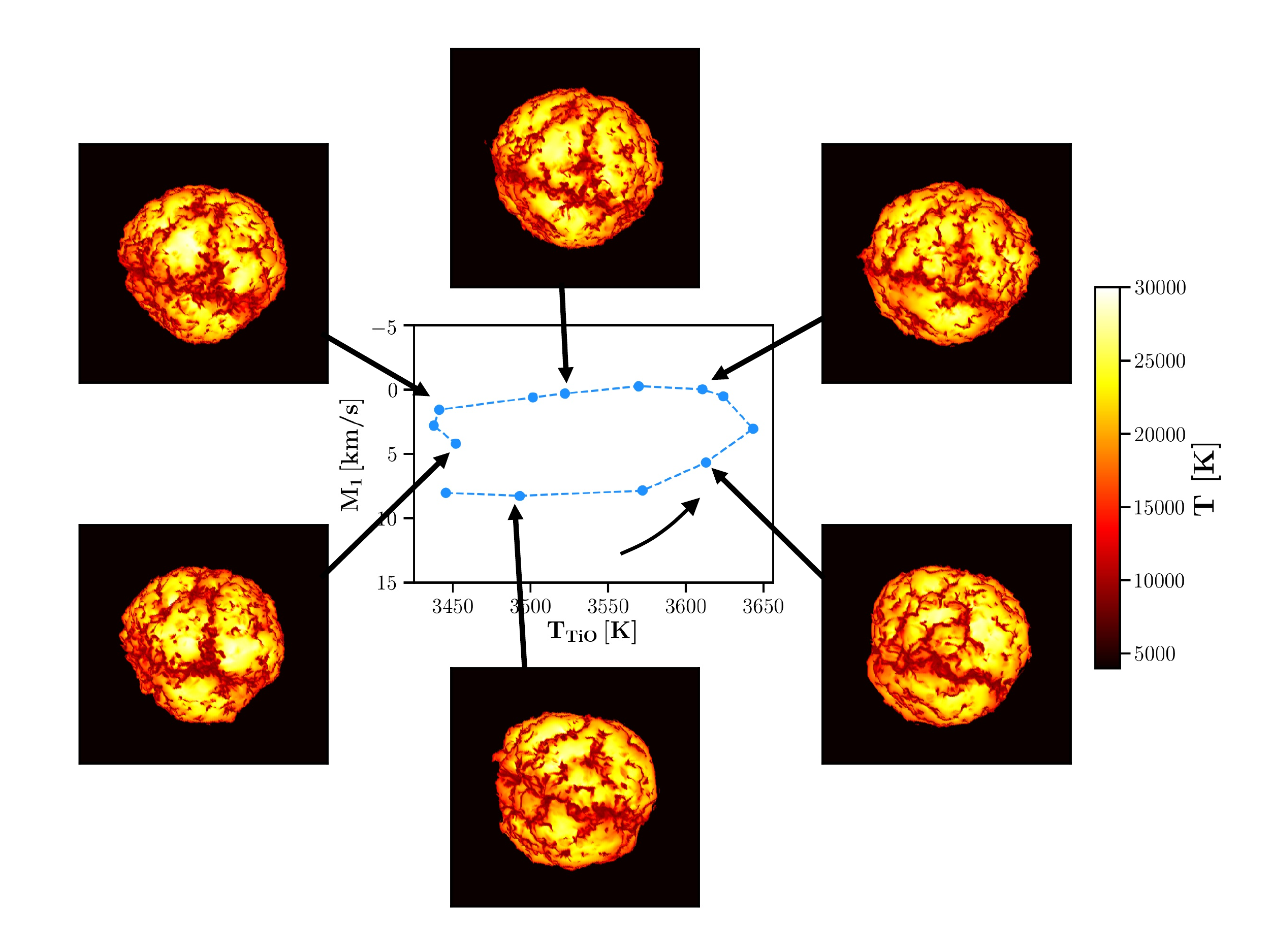}

      \caption{ Same as Fig.~\ref{maps_for_loops} for $\tau_0 = 20000$. Variations reflected by the hysteresis loop are not observed at this optical depth.
      }
         \label{maps_for_loops_largetau}
   \end{figure*}

According to Fig.~\ref{Hp_vs_masks}, the velocity and temperature maps of Fig.~\ref{maps_for_loops} describe the convective pattern located well above the bottom of the photosphere since the atmospheric layers probed by the tomographic masks are far up  the Rosseland radius marking the stellar surface. The corresponding pressure scale-height values are relatively small (reflecting small-scale surface granulation) with respect to those in the \textit{deep} convective zone (defined as the region below the Rosseland radius) where large convective cells are located (see Figs.~\ref{Fig:RV_map} and \ref{maps_for_loops_largetau}). An increase of the size of convective structures with increasing atmospheric depth is observed as well in convection simulations of main-sequence F-type stars \citep{2016ApJ...821L..17K}.

Figure~\ref{maps_for_loops_largetau} shows the velocity and temperature maps corresponding to the deep convection zone at $\tau_0 = 20000$ (see Fig.~\ref{Hp_vs_masks}) for the same 3D snapshots as in Fig.~\ref{maps_for_loops}. The large convective cells are clearly visible as bright granules on the temperature maps. Following the velocity maps, the matter is rising through the granules and falling in the intergranular lanes, thus in accordance with the classical convection scenario.

The rising material observed in Fig.~\ref{maps_for_loops} originates in the deep convective zone and an atmospheric shock develops higher up along the following steps. \citet{2017A&A...600A.137F} and \citet{2018A&A...619A..47L} illustrated this process in similar 3D RHD simulations: 

\begin{itemize}
    \item non-stationary convection (e.g., merging downdrafts or other localized events) produces a sound wave in the stellar interior;
    \item the sound wave travels through the star, until it hits the surface;
    \item at the surface, the wave is slowed down and compressed due to the drop in temperature and sound speed;
    \item in addition, the amplitude of the sound wave rises due to the decrease in density and, therefore, turns into a shock;
    \item a shock, then, propagates all the way from the stellar surface to the outer atmospheric layers (see Fig.~\ref{Fig:RV_map}).
\end{itemize}
While this happens for the Sun far out in the chromosphere, it happens 
for RSGs already in the photosphere, since all velocities and also Mach numbers are larger in RSGs \citep{2013EAS....60..137F}.
Therefore, the entire RSG spectrum and also the surface brightness are affected, and
   not only some exotic chromospheric emission lines as in the Sun.


The  link between the maps in  Figs.~\ref{maps_for_loops} and \ref{maps_for_loops_largetau} is best illustrated by Fig.~\ref{Fig:RV_map}, which presents radial-velocity maps in an equatorial plane for the snapshots along the hysteresis loop of mask C4 in Fig.~\ref{3D_hysteresis} (the online version of the paper displays an animated version of this figure). 
It reveals that convection is structured through the motions going all the way from the center of the star to the surface, separated by the downdrafts appearing as fingers on  Fig.~ \ref{Fig:RV_map}. That figure clearly reveals the structural change occurring between deep and surface convection.


%

\subsection{Convective and acoustic timescales}
\label{convective_lifetime}

As an important preamble to this section, one should realise that stationary convection cannot generate time-dependent effects like a hysteresis loop. Its origin must be looked for in time-dependent effects. These could originate from acoustic waves. These waves could disturb the top of the convective pattern, just below the continuum-forming region, and move to the line-forming regions where they modify the flow pattern, as discussed in Sect.~\ref{surface_vs_deep_convection}. 
In this section we investigate whether convective or acoustic timescales may account for the characteristic timescale of the  hysteresis loops in RSGs (Table~\ref{tab:mucep_loops}), which is of the same order as the observed photometric variations. As explained above (Sect.~\ref{surface_vs_deep_convection}), the surface features are probably not genuine convection, but are rather linked to shock waves originating in acoustic waves generated by the deep convection. Therefore, we shall consider both convective and acoustic (i.e., sound-crossing) timescales.

The sound velocity can be expressed as:
\begin{equation}
    v_s = \sqrt{\Gamma_1 \frac{\Re T_r}{\mu}} \enspace ,
\label{Eq.sound_speed}
\end{equation}
\noindent
where $\Gamma_1$ is the first adiabatic exponent ($\Gamma_1 = 5/3$ for a mono-atomic perfect gas), $T_r$ is the stellar temperature at radial distance $r$ from the stellar center, and $\mu$ = 1.3 $\rm g \, mol^{-1}$ for the atmosphere of a late-type star \citep{2011A&A...535A..22C}. 

A single convective timescale is much harder to define, as there is a whole range of them, related to the different convective spatial scales, all the way from the big convective cells to turbulent eddies. Nevertheless, to fix the ideas, one may use the mean lifetime of a turbulence element in the convective zone as expressed by the Mixing Length Theory (MLT) of convection 
\citep{1958ZA.....46..108B}:
\begin{equation}
t_{c} = \frac{\Lambda}{v_c} \enspace ,    
\label{Eq:t_conv}
\end{equation}
\noindent
where $\Lambda \simeq H_P$ is the mixing length (i.e., the vertical length travelled by a convective globule before dissolving in its environment) and $v_c$ is the velocity of the convective globule. 
Following \citet{2017use..book.....L}, the convective velocity can be derived from the relation:
\begin{equation}
v_c^3 = \frac{L_r}{6 \pi r^2} \Bigg( \frac{\rho_r \, \Re}{\mu}  \Bigg)^{-1} \frac{g_r \; \Lambda}{T_r} \enspace , 
\label{Eq:convective_velocity}    
\end{equation}
\noindent
where $T_r$, $L_r$, $\rho_r$ and $g_r$ are the temperature, stellar luminosity, density, and gravity at distance $r$ from the stellar center.

The sound and convective velocities are first derived for the deep interior ($r = R/2$, where $R$ is the stellar radius) and then for the surface ($r = R$), adopting the relevant physical quantities from a snapshot of the 3D simulation. The temperature $T_r$, density $\rho_r$, and pressure scale height $H_P(r)$ are computed at $r = R/2$ and $R$ by averaging the 3D snapshot variables over spherical shells \citep[as explained in][]{2009A&A...506.1351C}.
In the deep interior,  Eqs.~\ref{Eq.sound_speed} and~\ref{Eq:convective_velocity} predict $v_s$ = 33.8 km/s and $v_c = 15.2$ km/s, whereas in the outer layers  $v_s = 6.3$~km/s.

The convective velocity derived from Eq.~\ref{Eq:convective_velocity} above may be compared to the value directly extracted from the 3D snapshot. For each grid point of the considered 3D snapshot, we computed the convective velocity as $v_{c}^2 = (v_x^2 + v_y^2 + v_z^2)$,  where $v_x$, $v_y$ and $v_z$ are velocities along the $x$, $y$ and $z$ directions of the  simulation box. The $v_{c}$ velocity averaged in the range 0.45 -- 0.55~$R$ amounts to 12.3~km/s  and agrees well with the value  derived from Eq.~\ref{Eq:convective_velocity}.

In order to derive a convective timescale (amidst an extended spectrum of timescales, according to the above caveat) in the deep interior of the 3D simulation, we use Eq.~(\ref{Eq:t_conv}) with $\Lambda = H_P(r = R/2) \simeq 60$~R$_{\odot}$ (see Fig.~\ref{Hp_vs_masks}). The corresponding timescale is of the order of 30~d, i.e., much too short to account for those involved in the hysteresis loops (Table~\ref{tab:mucep_loops}). Further up in the atmosphere, the timescale would become even shorter since $H_P$ decreases outwards. Further down in the star, where large convective cells are located, the respective timescale would reach an order of years. Thus, the convective timescales cover a wide range from weeks to years, that would comprise the 324-day timescale of the simulated histeresis loop at some specific depth in the star. However, convection is not able to produce a particular, clearly observable feature with that precise timescale.

Instead, the acoustic timescale in the outer layers is a much better choice, assuming the vertical length travelled by a shock or acoustic wave to be equal to the distance between the stellar surface (at $r = R = 582$~ R$_{\odot}$ on Fig.~\ref{Hp_vs_masks}) and the atmospheric layer probed by mask C4 (at $r \simeq 800$~R$_{\odot}$ in the same figure). By adopting $v_s = 6.3$~km/s  in the outer layers as obtained above, the sound-crossing timescale amounts to 279~d,  close to the 324~d characteristic time of the hysteresis loop reported in Table~\ref{tab:mucep_loops}. This confirms our above guess that hysteresis loops must have an acoustic origin, although not fully independent of convection, since the acoustic wave modulates the surface flow keeping the memory of the underlying convective structure \citep{2017A&A...600A.137F}.

\section{Conclusions and future prospects}
\label{Sect:conclusions}

The present paper applies the tomographic method to the RSG star $\mu$ Cep in order to recover its line-of-sight velocity distribution, over the stellar disk, within different optical-depth slices and to relate it to the photometric variations. The observed velocity variations follow the photometric and temperature variations with a phase lag. This phase lag results in hysteresis loops in the temperature -- velocity plane which are characterized by timescales of a few hundred days, similar to the photometric ones. The same behavior was observed by \citet{2008AJ....135.1450G} for the RSG star Betelgeuse.

The hysteresis loop was also detected in the 3D RHD CO5BOLD simulation of a RSG star atmosphere. The qualitative agreement between the amplitudes of RV and temperature variations as well as of the timescales between observed and simulated hysteresis loops indicates that physical processes related to convection are responsible  for the few-hundred-day photometric variations in $\mu$ Cep and Betelgeuse. 

The timescales of observed and simulated hysteresis loops were compared to theoretical predictions of convective and sound-crossing timescales. 
It is found that the sound-crossing timescale in the outer layers is of the same order as the hysteresis-loop timescale. This suggests that hysteresis loops are linked to acoustic waves originating from a disturbance in the convective flow below the photosphere and propagating upwards to the surface layers where they modulate the convective energy flux.

Perspectives of this work include the application of the tomographic method to an extended sample of RSG stars in order to investigate whether the presence of  hysteresis loops is a common feature among them. This study is deferred to a forthcoming paper.

\begin{acknowledgements}

K.K acknowledges the support of a FRIA (FNRS) fellowship. S.V.E. thanks to Fondation ULB
for its support. This work is based on observations obtained with the HERMES spectrograph, which is supported  by  the  Fund  for  Scientific  Research  of  Flanders  (FWO),  Belgium,  the Research Council of K.U. Leuven, Belgium, the Fonds de la Recherche Scientifique (F.R.S.-FNRS), Belgium, the Royal Observatory of Belgium, the Observatoire de Gen{\`e}ve, Switzerland and the Th{\"u}ringer Landessternwarte Tautenburg, Germany. We acknowledge with thanks the variable star observations from the AAVSO International Database contributed by observers worldwide and used in this research.

\end{acknowledgements}

\bibliographystyle{aa}
\bibliography{bibliography}

\begin{appendix}

\section{Numerical resolution of 3D simulations}
\label{problem_with_numerical_resolution}

\begin{figure}[h]
   \centering

   \includegraphics[width=7cm]{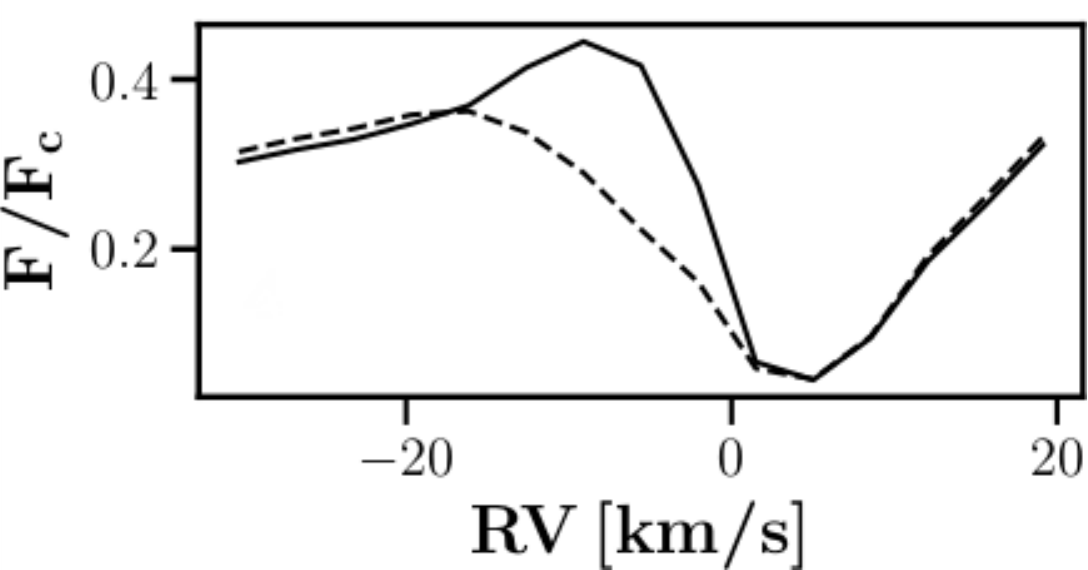}
  
      \caption{Continuum-normalized synthetic spectrum around $\lambda_0~=~4359.64$~{\AA} (around a line contributing to the mask C5) for a 3D snapshot. The wavelength axis was transformed into velocity. The solid line profile is produced using all the rays of the 3D snapshot while the dashed line shows the profile obtained by selecting only rays characterized by a good optical-depth sampling (see text).
      }
         \label{appendix_line_profiles}
   \end{figure}

As was already mentioned in \citet{2009A&A...506.1351C} and \citet{2018A&A...610A..29K}, some rays of the 3D simulation are characterized by large differences in the optical depth values between two adjacent grid points near the continuum-forming layers. This, in turn, causes uncertainties in the surface intensity computation: in the case of strong spectral lines, the surface intensity reaches values higher than that of the local continuum and produces spurious emission-like features in line profiles. For example, Fig.~\ref{appendix_line_profiles} shows continuum-normalized synthetic spectrum around the $\lambda_0 = 4359.64$~{\AA} line contributing to mask C5 (probing the outermost atmospheric layer) for a snapshot from the 3D simulation. It is characterized by an "emission" feature (with respect to the local continuum) in the blue-shifted wing of the spectral line.

In the meantime, we tried to resolve the issue associated with a poor optical-depth sampling by interpolating linearly the temperature (and the density) on a finer geometrical- or optical-depth grid along 3D snapshot rays. The first approach implied an oversampling of the snapshot box twice (resulting into a $801^3$ data cube); the corresponding variables (temperature, density, velocity) were interpolated at intermediate grid points. However, this approach allowed us to obtain only two times more data points in the required optical-depth region, but still lacking enough data for a reliable intensity computation (i.e. with no emission features). The second approach involved an oversampling of the optical depth region between $\log \tau =$ 0 and 2 by at least factor of five. However, since the true temperature (or density) profile in these regions is unknown, the linear interpolation could not produce a reliable intensity.  

   \begin{figure}
   \centering

   \includegraphics[width=8.5cm]{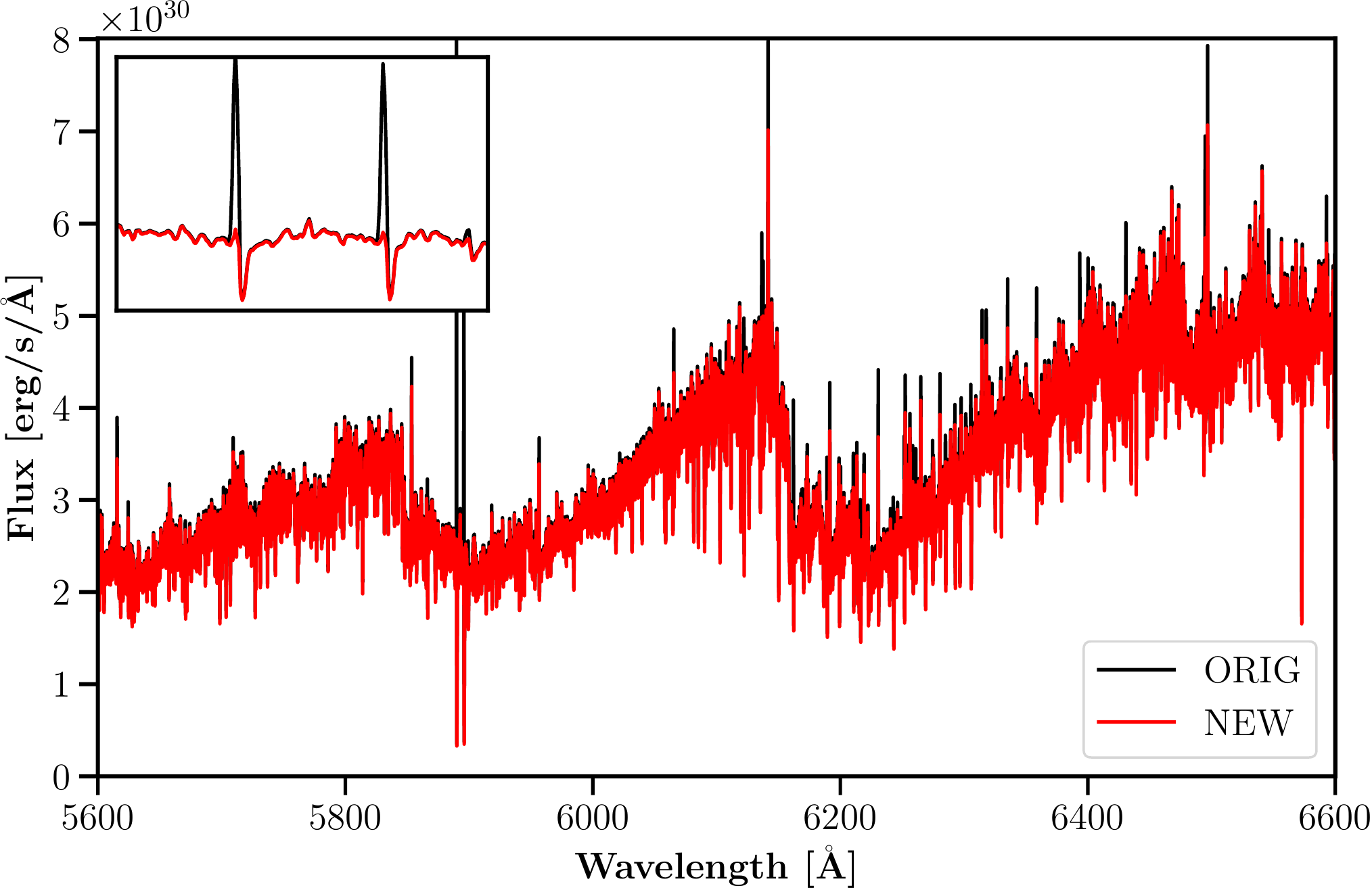}
  
      \caption{Synthetic spectra computed for a 3D snapshot with (red) and without (black) removing rays with poor optical-depth sampling. The inset shows a zoom between 5885 and 5900 $\AA$. As a result, most of emission features are suppressed without affecting the overall spectrum.
      }
         \label{appendix_spectra}
   \end{figure}

An optimal solution was nevertheless found, and it is based on computing the intensity only for rays characterized by a good optical-depth sampling near the continuum-forming layer. Several tests were performed in order to find an optimal number of points in the optical-depth range between 1 and 100, which are required to obtain the reliable integrated intensity. Our solution requires at least 5 points between optical-depth values of 1 and 100. It was applied to all 3D snapshots and allowed to keep at least 90 percent\footnote{Since not all the rays were used for the spectrum synthesis, the total flux was scaled by $1/f$, where $f$ is the fraction of rays used for the flux computation.} of rays per snapshot, thus not introducing significant effect on the overall spectrum (see Fig.~\ref{appendix_spectra}).  Moreover, the resulting line profile (displayed as the dashed line in Fig.~\ref{appendix_line_profiles}) does not show any artificial emission feature any more. 
The procedure described above was applied to all 3D snapshots analyzed in Sect.~\ref{Sect:3D_simulations}.

\section{Additional figures}

\begin{figure*}
\centering
   \includegraphics[width=15cm]{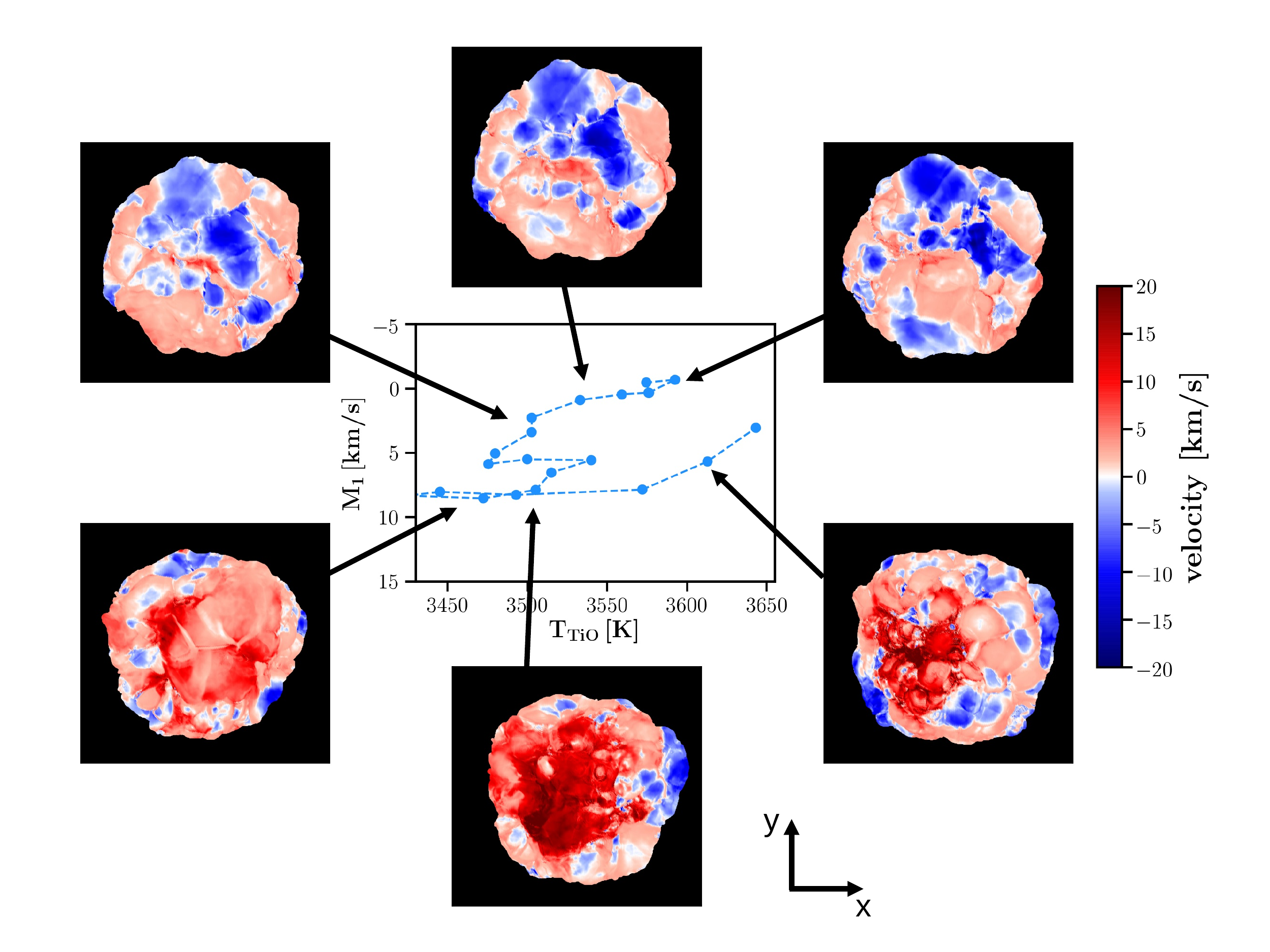}\\
   \includegraphics[width=15cm]{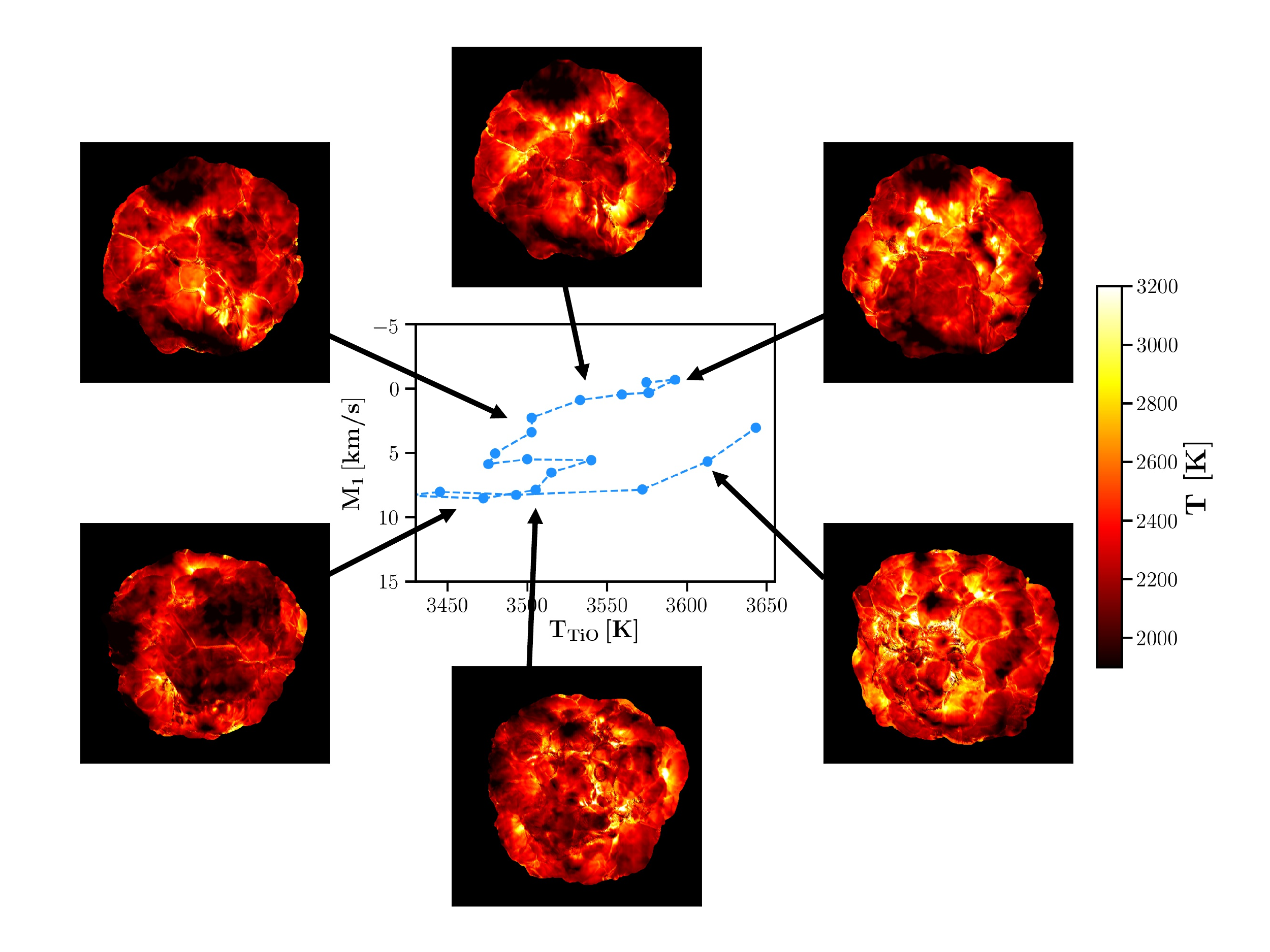}

      \caption{Same as Fig.~\ref{maps_for_loops} for the time range between days 1500 and 1950 (Fig.~\ref{simulated RVs 2}).
      }
         \label{maps_for_loops_NEW}
   \end{figure*}

\end{appendix}

\end{document}